%% file: ICC_Workshop.tex
\documentclass[conference]{IEEEtran}
\usepackage{amsmath,graphicx}
\usepackage{acronym}

\usepackage{tabularx, boldline, multirow}

\acrodef{RIS}{reconfigurable intelligent surface}
\acrodef{OFDM}{orthogonal frequency division multiplexing}
\acrodef{MIMO}{multiple-input-multiple-output}
\acrodef{SIMO}{single-input-multiple-output}
\acrodef{LoS}{line-of-sight}
\acrodef{AP}{access point}
\acrodef{CP}{carrier phase}
\acrodef{NCP}{no carrier phase}
\acrodef{NLoS}{non-line-of-sight}
\acrodef{SMC}{specular multipath component}
\acrodef{DMC}{dense multipath component}
\acrodef{SNR}{signal-to-noise ratio}
\acrodef{DNR}{dense-multipath-to-noise ratio}   %
\acrodef{SDNR}{signal-to-dense multipath-plus-noise ratio}
\acrodef{RS}{radio stripe}
\acrodef{SINR}{signal-to-interference-noise ratio}
\acrodef{BS}{base station}
\acrodef{UE}{user equipment}
\acrodef{NR}{new radio}
\acrodef{FR2}{frequency range 2}
\acrodef{DL}{downlink}
\acrodef{CPU}{central processing unit}
\acrodef{ISAC}{integrated sensing and communications}
\acrodef{UL}{uplink}
\acrodef{TDoA}{time-difference-of-arrival}
\acrodef{KPI}{key performance indicator}
\acrodef{AoA}{angle-of-arrival}
\acrodef{AoD}{angle-of-departure}
\acrodef{ULA}{uniform linear array}
\acrodef{ML}{maximum likelihood}
\acrodef{PEB}{position error bound}
\acrodef{CEB}{clock error bound}
\acrodef{RMSE}{root-mean-squared error}
\acrodef{FIM}{Fisher information matrix}
\acrodef{EFIM}[EFIM]{equivalent \ac{FIM}}   %
\acrodef{CRB}[CRLB]{Cram\'er-Rao lower bound}
\acrodef{LMR}{LoS-to-multipath ratio}
\acrodef{SRS}{sounding reference signal}
\acrodef{ILS}{iterative least squares}

\usepackage{amsmath,amssymb,epsfig,psfrag,cite,subfigure}
\include{macros}
\usepackage{graphicx}
\usepackage{epstopdf}

\usepackage{bm}

\usepackage{geometry}
\usepackage{tcolorbox}

\usepackage{accents}
\newcommand*{\dt}[1]{%
	\accentset{\mbox{\large .}}{#1}}

\allowdisplaybreaks

\usepackage{soul}

\usepackage{amsthm}

\usepackage{nicefrac}

\usepackage{lipsum,multicol}

\usepackage{algorithm}
\usepackage{algpseudocode}

\algdef{SE}[SUBALG]{Indent}{EndIndent}{}{\algorithmicend\ }%
\algtext*{Indent}
\algtext*{EndIndent}

\usepackage{enumitem}
\usepackage{mwe}    
\usepackage{epsfig,psfrag}
\usepackage{subfigure}
\usepackage{color}
\usepackage{url}
\usepackage{mathtools,xparse}

\usepackage{gensymb}

\usepackage[multiple]{footmisc}

\usepackage{balance}        	
\usepackage{siunitx}        		
\DeclareSIUnit{\SIdeg}{deg}	
\newlength{\plotWidth}		
\newlength{\plotHeight}		
  \usepackage{pgfplots}
  \pgfplotsset{compat=newest}
  \usetikzlibrary{plotmarks}
  \usetikzlibrary{arrows.meta,angles,quotes,3d,spy}
  \usepgfplotslibrary{patchplots}
  \usepackage{grffile}
  \usepackage{amsmath}
  \usepackage{tikzscale}		

\usepackage{xcolor}

\include{Other/commands}

\graphicspath{{./Figures/}}

\begin{document}
\bstctlcite{IEEEexample:BSTcontrol}

\title{Uplink Joint Positioning and Synchronization in Cell-Free Deployments with Radio Stripes}

\author{Alessio Fascista\IEEEauthorrefmark{1}, 
Benjamin J. B. Deutschmann\IEEEauthorrefmark{2}, 
Musa Furkan Keskin\IEEEauthorrefmark{3},  
Thomas Wilding\IEEEauthorrefmark{2}, \\ 
Angelo Coluccia\IEEEauthorrefmark{1},
Klaus Witrisal\IEEEauthorrefmark{2},
Erik Leitinger\IEEEauthorrefmark{2}, 
Gonzalo Seco-Granados\IEEEauthorrefmark{4},  
Henk Wymeersch\IEEEauthorrefmark{3}\\
\IEEEauthorrefmark{1}Universit{\`a} del Salento, Italy, 
\IEEEauthorrefmark{2}Graz University of Technology, Austria,\\
\IEEEauthorrefmark{3}Chalmers University of Technology, Sweden, 
\IEEEauthorrefmark{4}Universitat Autonoma de Barcelona, Spain}

\maketitle

\begin{abstract}
     Radio stripes (RSs) is an emerging technology in beyond 5G and 6G wireless networks to support the deployment of cell-free architectures. In this paper, we investigate the potential use of RSs to enable joint  positioning and synchronization in the uplink channel at sub-6 GHz bands. The considered scenario consists of a single-antenna user equipment (UE) that communicates with a network of multiple-antenna RSs distributed over a wide area. The UE is assumed to be unsynchronized to the RSs network, while individual RSs are time- and phase-synchronized. We formulate the problem of joint estimation of position, clock offset and phase offset of the UE and derive the corresponding maximum-likelihood (ML) estimator, both with and without exploiting carrier phase information. To gain fundamental insights into the achievable performance, we also conduct a Fisher information analysis and inspect the theoretical lower bounds numerically. Simulation results demonstrate that promising positioning and synchronization performance can be obtained in cell-free architectures supported by RSs, revealing at the same time the benefits of carrier phase exploitation through phase-synchronized RSs.

\end{abstract}

\section{Introduction}

Cell-free massive \ac{MIMO} has recently emerged as a promising technology for beyond 5G wireless networks to overcome challenges associated with conventional network-centric implementations, such as inter-cell interference and large intra-cell variations in data rate \cite{cellfree_2015,demir2021foundations,ubiquitous_cellFree_2019}. In user-centric cell-free architectures, each \ac{UE} communicates with a \ac{UE}-specific subset of widely distributed \acp{AP} that cooperatively serve it using phase-synchronized transmission/reception enabled by fronthaul links \cite{demir2021foundations}. Such an architecture not only improves communication metrics (i.e., more uniform coverage and better interference management), but also brings significant benefits for positioning and sensing \cite{nearfieldSense_TWC_2022,power_JRC_cellfree,JRC_cellfree_2022,nearfieldSense_Exp_2022}, which is an  opportunity for \ac{ISAC} in the cell-free context \cite{ISAC_2022}. In particular, phase-coherent processing with wide aperture enables exploitation of \textit{wavefront curvature} effects (i.e., near-field) \cite{nearfieldSense_TWC_2022,nearfieldSense_Exp_2022} and high-resolution \textit{\ac{CP}} information \cite{MIMO_radar_2008} to estimate \ac{UE}/object positions.

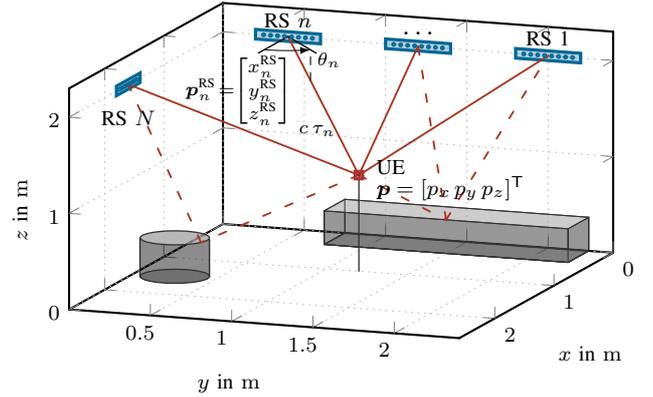
\begin{figure}
     \centering
     \setlength{\plotWidth}{\linewidth}
     \input{Figures/scenario.tex}\vspace{-3mm}
     \caption{General scenario of uplink UE positioning supported by a \acf{RS} network with $N$ stripes.}
     \label{fig:scenario}\vspace{-5mm}
 \end{figure}

Among cell-free implementation alternatives, the \acp{RS} technology holds great potential as a cost-efficient architecture for dense area deployments, such as stadiums and railway stations \cite{ubiquitous_cellFree_2019,seqRS_TCOM_2021}. \acp{RS}, also called RadioWeaves \cite{beamSync_RW,WPT_location_2022,RW_Indoor_2022}, consist of multiple antenna elements and processing units fitted inside the same cable, which can be easily deployed over a large area \cite{ubiquitous_cellFree_2019}. Serially connected \acp{RS} communicate with a \ac{CPU} via a shared bus that simultaneously provides synchronization and power supply \cite{ubiquitous_cellFree_2019}. From the viewpoint of positioning and sensing, synchronization of distributed arrays has been experimentally shown to improve \ac{AoA} estimation accuracy \cite{sync_AOA_TAP_2019}. Hence, phase-synchronized and distributed \acp{RS} offer a viable cell-free deployment solution to reap the benefits  in positioning and sensing.

Despite a considerable amount of research on the communications \cite{beamSync_RW,seqRS_TCOM_2021,RW_Indoor_2022,RS_conf_2021}, wireless power transfer \cite{RS_WPT_TWC_2022,WPT_location_2022}, and sensing \cite{nearfieldSense_Exp_2022,nearfieldSense_TWC_2022} aspects of \acp{RS}, no studies have investigated the potential of positioning aided by widely distributed \acp{RS} at sub-6 GHz bands. Recent work investigated coherent localization in distributed millimeter-wave (mmWave) massive \ac{MIMO} systems \cite{coherentDistMIMO_loc_2019,coherentDistMIMO_loc_2018,coherentDistMIMO_loc_CRB_2021}. However, two major differences exist between sub-6 GHz and mmWave operation. First, ensuring phase synchronization and calibration among distributed arrays at mmWave bands is extremely challenging since hardware imperfections (e.g., phase noise and frequency errors) become more severe as the carrier frequency increases \cite[Sec.~6.2.1]{hexax_d22}. Second, unlike the sparse nature of mmWave propagation, sub-6 GHz channels involve \acp{DMC} \cite{richter2005estimation,WPT_location_2022} that should be incorporated into signal modeling as a disturbance with certain statistical characterization. Therefore, the question remains unanswered as to \textit{under what conditions and to what extent phase synchronization and accompanying phase-coherent processing can improve positioning in a widely distributed \acp{RS} network at sub-6 GHz}. To fill this knowledge gap, this paper addresses the problem of uplink positioning and synchronization of a \ac{UE} supported by a network of \acp{RS} geographically distributed over a large area in a cell-free deployment scenario. In contrast to co-located massive \ac{MIMO} based localization (e.g., \cite{mMIMO_CPP_TWC_2019}), distributed \ac{MIMO} setups with phase-coherent processing enable exploitation of spherical wavefront through near-field conditions and \ac{CP} information to obtain high-resolution location estimates. The main contributions of the paper are as follows:
\begin{itemize}
    \item We investigate the problem of uplink joint positioning and synchronization of a \ac{UE} with the  emerging technology of distributed \acp{RS}, considering the distinctive properties of sub-6 GHz operation, including phase and time synchronization capability \cite{REINDEER_D2_1}, dense multipath environment \cite{WPT_location_2022} and \ac{CP} exploitation. 
    \item We derive the \ac{ML} estimators and the corresponding \acp{CRB} on position, clock and phase offset estimation, both with and without exploiting \ac{CP} information.
    \item We carry out extensive simulation analysis to showcase the impact of various system parameters (i.e., bandwidth, aperture size, \ac{SDNR}, presence/absence of phase synchronization) on positioning and clock offset estimation accuracy, offering valuable insights into practical \ac{RS} deployments towards 6G networks.
\end{itemize}

\textit{Notations:} $\toep(\xx, \xx^\hermit)$ denotes a Hermitian Toeplitz matrix with first column $\xx$ and first row $\xx^\hermit$. $\veccinv{\cdot}$ reshapes a vector into an $M \times K$ matrix. $[\bm{x}]_i$ denotes the $i$-th element of  the vector $\bm{x}$ and $[\bm{X}]_{i,j}$ the element of row $i$ and column $j$ of the matrix $\bm{X}$. The vector $\bm{1}_n$ is $1$ at the $n$-th entry and $0$ elsewhere. $\realp{x}$ denotes the real part of $x$.

\section{System Model and Problem Formulation}
In this section, we describe the uplink positioning scenario with \acp{RS} acting as receivers and a single \ac{UE} in the role of transmitter, illustrate the signal model at the \acp{RS} and formulate the joint positioning and synchronization problem.

\subsection{Uplink Positioning Scenario with Radio Stripes}
Consider a \ac{SIMO} system with a \acp{RS} network composed of $N$ stripes, each consisting of $M$ antennas, and a single \ac{UE} equipped with a single antenna and communicating with the network through the \ac{UL} channel \cite{seqRS_TCOM_2021,RS_WPT_TWC_2022}. The \acp{RS} network is assumed to have perfect \textit{phase synchronization} among individual stripes, thereby effectively turning it into a large multiple-antenna access point \cite[Sec.~3.1]{ubiquitous_cellFree_2019}, while the \ac{UE} has unknown \textit{phase offset} $\deltaphi$ and unknown \textit{clock offset} $\deltatau$ with respect to the \acp{RS} network. The wavefront of the signal transmitted by the \ac{UE} is assumed planar over each individual RS (i.e., individual \acp{RS} lie in the far-field of the \ac{UE}), but no longer planar over the \acp{RS} network seen as a whole due to a large aperture distributed over a wide area. The \acp{RS} are deployed around a rectangle 
located at a specific height from the floor level, as depicted in Fig.~\ref{fig:scenario}. Each individual \ac{RS} is placed at a known position $\pprs_n = [\xrs_n \ \yrs_n \ \zrs_n]^\trp$ with known orientation $\beta_n$ around the $z$-axis,  measured counter-clockwise from the $x$-axis\footnote{The orientation of each RS is defined by a single angle representing the rotation around the $z$-axis, meaning that the RSs are completely aligned with the $x-y$ plane of the coordinate system.}, while the \ac{UE} is located at an unknown position $\pp = [p_x \ p_y \ p_z]^\trp$.

\subsection{Signal and Channel Models}
For \ac{UL} communications, the \ac{UE} transmits \ac{OFDM} pilots $\sss = [ s_0 \ \cdots \ s_{K-1} ]^\trp \in \complexset{K}{1}$ over $K$ subcarriers with subcarrier spacing $\deltaf$, e.g., \ac{SRS} for 5G \ac{NR} \ac{UL} positioning \cite{3GPP_1810532}. 
Assuming a quasi-static block fading condition, the UL received signal at the $n$-th RS over subcarrier $k$ can be written as \cite{mMIMO_CPP_TWC_2019,WPT_location_2022}
\begin{align}\label{eq_ynk}
    \yy_{n,k} = \hh_{n,k} s_k + \wwdmc_{n,k} s_k + \zz_{n,k} \in \complexset{M}{1} ~,
\end{align}
where $\hh_{n,k} \in \complexset{M}{1}$ denotes the deterministic channel components including the \ac{LoS} path and the possible \ac{NLoS} contributions originating from dominant reflections from large objects in the surrounding environment, $\wwdmc_{n,k} \in \complexset{M}{1}$ represents the contribution from \acp{DMC}, and $\zz_{n,k} \in \complexset{M}{1} $ denotes circularly symmetric complex Gaussian noise with $\zz_{n,k} \sim \mtCN(\boldzero_M, \allowbreak \sigma^2 \Imatrix_M )$, with $\sigma^2$ denoting the noise power. In this work, we consider the simplified scenario in which the \ac{LoS} path over each UE-RS link is assumed to be dominant compared to the additional \ac{NLoS} paths, 
giving rise to the following geometric channel model
\cite{mMIMO_CPP_TWC_2019,WPT_location_2022,RS_WPT_TWC_2022}
\begin{equation}\label{eq_SMC_channel}
    \hh_{n,k} =  \alpha_{n} e^{j\phi_{n}} \aaa(\theta_{n}) e^{-j 2 \pi k \deltaf \tauu_{n}} ~,
\end{equation}
where
\begin{itemize}
    \item $\alpha_{n} \in \realsetone{}$ is the large-scale fading amplitude coefficient. 
    \item $\phi_{n}$ is the phase term involving the effects of one-way signal propagation (related to the UE position $\pp$) and phase offset between the \acp{RS} network and  \ac{UE}, given by
    \begin{equation} \label{eq_phin}
        \phi_{n} = -2\pi\fc \tau_{n}  + \deltaphi    ~,
    \end{equation}
    with
    \begin{equation}\label{eq_taun}
    \tau_{n} = \frac{1}{c} \|\pp - \pprs_n  \|.
    \end{equation}
    \item $\tauu_{n}$ is the pseudo-delay including the effect of one-way propagation and the clock offset of the UE, namely
\begin{align} \label{eq_tau_pseudo}
    \tauu_{n} = \tau_{n} + \deltatau ~.
\end{align}
    \item $\aaa(\theta_{n}) \in \complexset{M}{1}$ is the RS array response to a signal impinging with \ac{AoA} $\theta_{n}$ (azimuth angle relative to the boresight of the $n$-th RS antenna array). Without loss of generality, we assume that each \ac{RS} is equipped with a \ac{ULA} with antenna element spacing\footnote{For \ac{RS} deployments, the element spacing can be larger than the standard half-wavelength spacing to increase spatial resolution \cite{RW_Indoor_2022}, \cite[Ch.~2.2.4]{REINDEER_D3_1}.} $d$, so that the array response vector takes the form
\begin{equation}
\aaa(\theta) \triangleq \left[e^{-j \frac{2\pi}{\lambda}d \left(\frac{M-1}{2} \right) \sin \theta} \ \cdots \ e^{j\frac{2\pi}{\lambda}d \left(\frac{M-1}{2} \right) \sin \theta}\right]^\trp~
\end{equation}
with $\lambda = c/\fc$, $f_c$ and $c$ denoting the wavelength, carrier frequency and speed of propagation, respectively. The \ac{AoA} $\theta_{n}$ relates the known position and orientation of the $n$-th RS and the unknown UE position according to
\begin{equation} \label{eq_thetanl}
    \theta_{n} = \frac{\pi}{2} - \mathrm{atan2}\left( [\pps_{n}]_2 , [\pps_{n}]_1 \right) ~,
\end{equation}
where $\mathrm{atan2}(y,x)$ denotes the four-quadrant arc tangent function and
\begin{equation} \label{eq_ppsnl}
    \pps_{n} =  \mmb^{-1}(\beta_n)(\pp - \pprs_n) 
\end{equation}
is the UE position in the local reference frame of the $n$-th RS, and $\mmb(\beta)$ is the rotation matrix around the $z$-axis.

\end{itemize}

\subsection{Dense Multipath Components}
The dense multipath term in \eqref{eq_ynk} can be modeled as a stochastic component with  distribution \cite{DMC_TAP_2014,mMIMO_CPP_TWC_2019}
\begin{align} \label{eq_dmc_stat}
    \wwdmc_n \sim \mtCN(\boldzero_{MK}, \allowbreak \rrbdmc(\etadmc) ) ~,
\end{align}
with $\wwdmc_n \triangleq [ (\wwdmc_{n,0})^\trp \ \cdots \ (\wwdmc_{n,K-1})^\trp ]^\trp \in \complexset{MK}{1}$
denotes the \ac{DMC} observed in the spatial-frequency domain, and $ \rrbdmc  \in \complexset{MK}{MK}$ is the spatial-frequency covariance matrix of the \ac{DMC}. Assuming spatially white \ac{DMC} and the Kronecker separability of the spatial and frequency domains (i.e., uncorrelated scattering between angle and delay domains), $\rrbdmc$ can be written as \cite[Eq.~(2.69)]{richter2005estimation}, \cite{DMC_TAP_2014,mMIMO_CPP_TWC_2019}
\begin{align} \label{eq_dmc_cov}
     \rrbdmc(\etadmc) = \rrb_f(\etadmc) \otimes \Imatrix_M  ~,
\end{align}
where $\rrb_f(\etadmc) \in \complexset{K}{K}$ is the frequency domain covariance matrix with a Toeplitz structure
\begin{align} \label{eq_rf}
    \rrb_f(\etadmc) = \toep( \kappab(\etadmc), \kappab(\etadmc)^\hermit  ) ~.
\end{align}
In \eqref{eq_rf}, $\etadmc = [ \alpha_d \ \beta_d \ \tau_d ]^\trp$ is the DMC parameter vector consisting of the peak power $\alpha_d$, the normalized coherence bandwidth $\beta_d$ and the normalized onset time $\tau_d$, and $\kappab(\etadmc) \in \complexset{K}{1}$ represents the sampled version of the DMC power spectral density \cite[Eq.~(2.61)]{richter2005estimation}
\begin{align}
    \psi_{\DMC}(f) = \frac{\alpha_d}{ \beta_d + j 2 \pi f } e^{-j 2 \pi f \tau_d} ~.
\end{align}

\subsection{Spatial-Frequency Observations at Radio Stripes}
Aggregating the received signals in \eqref{eq_ynk} over $K$ subcarriers and using the geometric model in \eqref{eq_SMC_channel} and the DMC model in \eqref{eq_dmc_stat}, the spatial-frequency observation matrix at the $n$-th \ac{RS} is
\begin{align} \nonumber
    \yyb_n &\triangleq [\yy_{n,0} \ \cdots \ \yy_{n,K-1}  ] \in \complexset{M}{K}
    \\ \label{eq_yy}
    &= \alpha_{n} e^{j\phi_{n}} \aaa(\theta_{n}) ( \bb(\tauu_{n}) \odot \sss )^\trp + \wwb_n ~,
\end{align}
where 
\begin{align}
    \bb(\tau) &\triangleq   \left[ 1 \ e^{-j 2 \pi \deltaf \tau} \ \cdots \ e^{-j 2 \pi (K-1) \deltaf  \tau} \right]^\trp \in \complexset{K}{1} 
\end{align}
is the frequency domain steering vector, and
\begin{align} \label{eq_wwb}
    \wwb_n = \wwbdmc_n \odot \boldone \sss^\trp + \zzb_n \in \complexset{M}{K}
\end{align}
represents the disturbance term consisting of the \acp{DMC} and additive white noise, with $\wwbdmc_n = \veccinv{\wwdmc_n} \in \complexset{M}{K} $ and $\zzb_n = [\zz_{n,0} \ \cdots \zz_{n,K-1}] \in \complexset{M}{K}$. After simple manipulations (details omitted due to lack of space), it can be shown that $\wwb_n$ has the distribution
\begin{align}
    \vecc{ \wwb_n } \sim \mtCN(\boldzero_{MK}, \allowbreak \rrb(\etadmc, \sigma^2) ) ~,
\end{align}
where 
\begin{align} \label{eq_r_dmc}
    \rrb(\etadmc, \sigma^2) = \left( \rrb_f(\etadmc) \odot \sss \sss^\hermit \right) \otimes \Imatrix_M + \sigma^2 \Imatrix_{MK} ~.
\end{align}
Different methods have been proposed to estimate the DMC parameters. For instance, in \cite[Sec.~6.1.8]{richter2005estimation} authors provide a suitable method for finding suboptimal though  accurate estimates of the DMC parameters starting from  covariance matrix estimates (obtained from the observed data) and exploiting the peculiar Toeplitz structure of $\rrb(\etadmc, \sigma^2)$. A similar method is adopted in \cite[Sec.~III-C1]{mMIMO_CPP_TWC_2019} to find good estimates of the DMC parameters and reconstruct an estimate of $\rrb(\etadmc, \sigma^2)$. In the following, to decouple the less investigated problem of joint localization and synchronization of a UE supported by a network of \acp{RS} from the more understood problem of estimating DMC parameters, we assume that a preliminary \emph{calibration} phase has been performed to estimate $\rrb(\etadmc, \sigma^2)$, by resorting to one of the methods presented in \cite{mMIMO_CPP_TWC_2019}, \cite{{richter2005estimation}}.

\subsection{Problem Formulation}\label{sec_prob_form}
In the considered \ac{UL} communication scenario supported by a network of \acp{RS}, our goal is to \emph{estimate} the position of the UE and to \emph{synchronize} its clock and phase to the \acp{RS} network. To ease the exposition and without loss of generality,we assume that the height of the UE ($p_z$ coordinate) is assigned and the ultimate positioning problem then consists in retrieving the $(p_x, p_y)$ coordinates, i.e., locating the UE in the $x-y$ plane.\footnote{Note that the $z$-component of the UE position can be determined using ULAs and delay information, though with a lower accuracy due to the absence of angular elevation information. The considered scenario can be easily extended to provide accurate 3D localization by using 2D planar arrays (e.g., rectangular) in place of ULA-RSs and by including the estimation elevation angles in the estimation problem. Note that no theoretical issues will arise when extending the proposed estimation approaches to 3D localization, except for a higher computational complexity due to the increased dimension of the estimation problem.}
Accordingly, $\pptwoD = [p_x \ p_y]^\mathsf{T}$ is the 2D UE position vector. 

Given the observations $\{\yyb_n\}_{n=0}^{N-1}$ in \eqref{eq_yy} collected from all \acp{RS}, the problem of interest is to estimate the UE position $\pptwoD$, its clock offset $\deltatau$ and its phase offset $\deltaphi$. The unknown parameter vector for this estimation problem is defined as
\begin{equation} \label{eq_etab}
    \etab    = [ \pptwoD^\trp \ \deltatau  \ \deltaphi \  \alphabar^\trp \     ]^\trp \in \realset{(N+4)}{1} ~,
\end{equation}
where $\alphabar \triangleq [ \alpha_0 \ \cdots \ \alpha_{N-1} ]^\trp \in \realset{N}{1}$.

\section{Joint Uplink Positioning and Synchronization}\label{sec::design_estimators}
In this section, we derive novel algorithms based on the \ac{ML} theory to solve the joint positioning and synchronization problem formulated in Sec.~\ref{sec_prob_form}.


\subsection{Joint Direct Positioning and Synchronization}\label{sec_dp_sc1}
Leveraging the \ac{ML} rationale, we formulate the positioning and synchronization problem as a \emph{direct} joint estimation problem where the sought $\pptwoD$, $\delta_\tau$, and $\delta_\phi$ parameters are directly inferred from the raw observations collected at all \acp{RS} as
\begin{align}\label{eq_direct_pos_sc1}
         \etabhat^{\text{\tiny ML}} = \arg \max_{\etab} ~ p(\{\yyb_n\}_{n=0}^{N-1} ~ \lvert ~ \etab ) ~.
    \end{align}
Assuming independent realizations of the disturbance component $\wwb_n$ in \eqref{eq_yy} across the \acp{RS}, the log-likelihood version of the objective function in \eqref{eq_direct_pos_sc1} can be written as
\begin{align} \label{eq_direct_lik}
    \log p(\{\yyb_n\}_{n=0}^{N-1} ~ \lvert ~ \etab ) = \sum_{n=0}^{N-1} \log p(\yyb_n ~ \lvert ~ \etab ) ~,
\end{align}
where 
\begin{align}\nonumber
    \log p(\yyb_n ~ \lvert ~ \etab ) &= - \normbig{ \rrb^{-1/2} \big[ \yy_n - \alpha_n e^{j \phi_n}  \cc(\theta_n, \tauu_n) \big] }_2^2 
    \\
    & ~~ -MK \log \pi - \log \det \rrb ~,
\end{align}
$\cc(\theta, \tau) \triangleq  ( \bb(\tau) \odot \sss ) \otimes \aaa(\theta) \in \complexset{MK}{1}$, $\yy_n \triangleq \vecc{\yyb_n} \in \complexset{MK}{1}$ and $\rrb$ is defined in \eqref{eq_r_dmc} (for conciseness, we omit the dependencies on $\etadmc$ and $\sigma^2$). Neglecting irrelevant constant terms in \eqref{eq_direct_lik}, the problem in \eqref{eq_direct_pos_sc1} becomes
\begin{align}\label{eq_direct_pos_sc1_2}
         \etabhat^{\text{\tiny ML}} = \arg \min_{\etab} ~ \llr_{\text{\tiny ML}}(\etab)  ~,
    \end{align}
where 
\begin{align} \label{eq_direct_pos_lik}
    \llr_{\text{\tiny ML}}(\etab) \triangleq \sum_{n=0}^{N-1} \normbig{ \yyp_n - \alpha_n e^{j \phi_n}  \ccp(\theta_n, \tauu_n) }_2^2 ~,
\end{align}
$\yyp_n \triangleq \rrb^{-1/2} \yy_n$ and $\ccp(\theta, \tau) \triangleq \rrb^{-1/2} \cc(\theta, \tau)$. To tackle the \ac{ML} estimation problem, we first notice that the amplitudes $\alpha_n$ in \eqref{eq_direct_pos_sc1_2} can be estimated in closed-form on a per-\ac{RS} basis as a function of the remaining parameters belonging to the respective \ac{RS} as
\begin{align}\label{eq_alphahat_n}
    \alphahat^{\text{\tiny ML}}_n & \!=\! \frac{\realp{ (e^{j \phi_n} \ccp(\theta_n, \tauu_n))^\hermit \yyp_n  } }{ \norm{e^{j \phi_n} \ccp(\theta_n, \tauu_n)}_2^2 } \!=\! \frac{\realp{ (e^{j \phi_n} \ccp(\theta_n, \tauu_n))^\hermit \yyp_n  } }{ \norm{\ccp(\theta_n, \tauu_n)}_2^2 }\nonumber \\
    & =\frac{\left((e^{j \phi_n} \ccp(\theta_n, \tauu_n))^\hermit \yyp_n 
    \!+\! 
    (e^{j \phi_n} (\yyp_n)^\hermit \ccp(\theta_n, \tauu_n)) 
    \right) }{2\norm{\ccp(\theta_n, \tauu_n)}_2^2}.
\end{align}
Plugging \eqref{eq_alphahat_n} back into \eqref{eq_direct_pos_lik}, utilizing \eqref{eq_phin} and dropping the dependency on $\alphabar$, we obtain the compressed log-likelihood
\begin{align} \label{eq_direct_pos_sc1_3}
         \llr_{\text{\tiny ML}}(\pptwoD, \deltatau, \deltaphi)  \!\!&=\!\! \sum_{n=0}^{N-1} \normbig{ \overbrace{\yyp_n  - \frac{(\ccp(\theta_n, \tauu_n))^\hermit \yyp_n}{2 \norm{\ccp(\theta_n, \tauu_n)}_2^2} \ccp(\theta_n, \tauu_n)}^{\triangleq \yybr_n(\theta_n, \tauu_n)} \nonumber
         \\ 
         &~~~~~~~~- \frac{e^{j 2 \phi_n} (\yyp_n)^\hermit \ccp(\theta_n, \tauu_n)}{2 \norm{\ccp(\theta_n, \tauu_n)}_2^2}  \ccp(\theta_n, \tauu_n) }_2^2  ~, \nonumber
         \\ 
         &=\!\! \sum_{n=0}^{N-1} \normbig{ \yybr_n(\theta_n, \tauu_n) - e^{j 2 \deltaphi} \ccbr_n(\theta_n, \tauu_n, \tau_n) }_2^2 ~,
\end{align}
where
\begin{align}
    \ccbr_n(\theta_n, \tauu_n, \tau_n) \triangleq \frac{e^{-j 4 \pi \fc \tau_n} (\yyp_n)^\hermit \ccp(\theta_n, \tauu_n)}{2 \norm{\ccp(\theta_n, \tauu_n)}_2^2}  \ccp(\theta_n, \tauu_n) ~.
\end{align}

It is not difficult to show that also the phase offset $\deltaphi$ can be estimated in closed-form as a function of the remaining parameters. More specifically, the optimal $\deltaphi$ minimizing \eqref{eq_direct_pos_sc1_3} can be readily obtained as
\begin{align} \label{eq_deltaphihat}
    \deltaphihat^{\text{\tiny ML}} =  \frac{ \angle \big( \sum_{n=0}^{N-1} \ccbr_n^\hermit(\theta_n, \tauu_n, \tau_n) \yybr_n(\theta_n, \tauu_n) \big)  }{2} + A \pi ~,
\end{align}
where $A \in \intset$ is introduced to account for possible integer ambiguities in phase estimation. Inserting \eqref{eq_deltaphihat} into \eqref{eq_direct_pos_sc1_3} yields
\begin{align} \label{eq_direct_pos_sc1_4}
         \llr_{\text{\tiny ML}}(\pptwoD, \deltatau) 
        &= \sum_{n=0}^{N-1} \big( \norm{ \yybr_n(\theta_n, \tauu_n) }_2^2 + \norm{\ccbr_n(\theta_n, \tauu_n, \tau_n) }_2^2 \big) \nonumber \\ 
         &~~~~- 2 \absbigg{  \sum_{n=0}^{N-1} \ccbr_n^\hermit(\theta_n, \tauu_n, \tau_n) \yybr_n(\theta_n, \tauu_n)} ~,
\end{align}
where the direct geometric relation between the position parameters $\pp = [\pptwoD \ p_z]^\mathsf{T}$, $ \deltatau$ and the per-RS channel parameters $\{\theta_n, \tauu_n, \tau_n\}_{n=0}^{N-1}$ is specified in \eqref{eq_taun}, \eqref{eq_tau_pseudo}, \eqref{eq_thetanl} and \eqref{eq_ppsnl}. The final expression of the direct ML estimator than becomes
\begin{align}\label{eq_direct_pos_final}
         [\hat{\pp}_{\text{\tiny 2D}}^{\text{\tiny ML}} \ \hat{\delta}_\tau^{\text{\tiny ML}}] = \arg \min_{\pptwoD,\delta_\tau} ~\llr_{\text{\tiny ML}}(\pptwoD, \deltatau)   ~,
    \end{align}
where, interestingly, the dependencies left are only upon the parameters of interest $\pptwoD$ and $\delta_\tau$. In principle, solving \eqref{eq_direct_pos_final} would require a
joint optimization over the continuous support defined by the three parameters $\pptwoD$ and $\delta_\tau$, which unfortunately is not feasible in closed-form. A more practical way to tackle \eqref{eq_direct_pos_final} consists in griding the parameters support and then employing an exhaustive, but computationally demanding 3D grid search. In the following Proposition, we illustrate an alternative approach to obtain a low-complexity estimate $\hat{\delta}_\tau$.

\begin{proposition}
    A coarse estimate of $\deltatau$ can be obtained by using a multilateration approach based on a low-complexity \ac{ILS} procedure \cite{ILS}. Specifically, by defining $\bar{\bm{Y}}_n = \mathrm{IFFT}(\yyb_n^\mathsf{T})$ as the IFFT-transformed observations over $N_F$ points for each UE-RS link, we first perform a noncoherent integration across the spatial domain  (i.e., samples over the $M$ antennas at each \ac{RS}) and seek for the index of the maximum element in the cost function
    $$
    \hat{q} = \arg \max_{q} \left[\sum_{m=1}^M |[\bar{\bm{Y}}_n]_{[q,m]}|^2 : 0 \leq q \leq N_F -1\right]
    $$
 with $|[\bar{\bm{Y}}_n]_{[q,m]}|$ denoting the absolute value of the $(q,m)$-th entry of $\bar{\bm{Y}}_n$. Accordingly, a coarse estimate of the pseudo-delay $\tauu_{n}$ can be obtained by mapping the index $\hat{q}$ with the corresponding IFFT bin as
 $
 \widehat{\tauu}_{n} = {\hat{q}}/{(N_F \Delta f)}.
 $
 Once the set of pseudo-delays $\{\widehat{\tauu}_{n}\}_{n=1}^N$ for all UE-RS links has been estimated, it can be used to set up a system of equations expressed as a function of the parameter of interest $\delta_\tau$ (in addition to $\pp$) according to \eqref{eq_tau_pseudo}. This over-determined system (for $N > 3$) can be efficiently solved by adopting an ILS procedure as detailed in \cite[Sec.~3.1]{JSAN_ILS}. We refer to the obtained clock offset estimate as $\hat{\delta}^{\text{\tiny ILS}}_\tau$.
\end{proposition}

The estimate $\hat{\delta}^{\text{\tiny ILS}}_\tau$ can be  plugged back into \eqref{eq_direct_pos_sc1_4} to reduce the cost involved in the optimization from 3D to 2D. Solving \eqref{eq_direct_pos_final} for a fixed $\delta_\tau = \hat{\delta}^{\text{\tiny ILS}}_\tau$ allows us to obtain also an initial estimate of $\pptwoD$ which, given its strong dependency on the ILS estimation accuracy, will be denoted for convenience as $\hat{\pp}_{\text{\tiny 2D}}^{\text{\tiny ILS}}$. The actual ML estimates that we retain can be then obtained using a further low-complexity refinement step where we solve \eqref{eq_direct_pos_final} for $(\pptwoD, \delta_\tau)$ jointly by means of a low-complexity iterative optimization (e.g., Nelder-Mead algorithm), using the suboptimal estimates $\hat{\pp}^{\text{\tiny ILS}}_{\text{\tiny 2D}}$ and $\hat{\delta}^{\text{\tiny ILS}}_\tau$  as initialization.

\subsection{Joint Direct Positioning and Synchronization without Exploiting Carrier Phase Information}
In this section, we derive an alternative version of the direct  estimator in \eqref{eq_direct_pos_sc1} where we do not exploit the \ac{CP} information in \eqref{eq_phin}. Our aim is to explore the impact of phase synchronization among the \acp{RS}, represented by the parameter $\deltaphi$ in \eqref{eq_direct_pos_sc1}, on localization accuracy (i.e., to study if accuracy degrades when $\phi_n$ is assumed to be an unknown parameter that has no relation to the geometry through $\tau_n$). In this case, by treating the channel amplitudes as unstructured complex entities $\gamma_n = \alpha_n e^{j \phi_n} \  \forall n$, the cost function in \eqref{eq_direct_pos_lik} becomes
\begin{align} \label{eq_direct_pos_lik_alt}
    \llr_{\text{\tiny ML-NCP}}(\etab) \triangleq \sum_{n=0}^{N-1} \normbig{ \yyp_n - \gamma_n \ccp(\theta_n, \tauu_n) }_2^2 ~,
\end{align}
where $\gamma_n \in \complexsett$, the label ``ML-NCP'' denotes \ac{NCP}, and the position-domain unknown parameter vector becomes
\begin{align}
    \etab    &= [ \pptwoD^\trp \ \deltatau  \ \realp{\gammab}^\trp \ \imp{\gammab}^\trp     ]^\trp \in \realset{(2N+3)}{1} ~,
\end{align}
with $\gammab \triangleq [\gamma_0 \ \cdots \ \gamma_{N-1}]^\trp$. In \eqref{eq_direct_pos_lik_alt}, the estimates of the complex gains $\gamma_n$, $n=1,\ldots,N$, can be obtained as
\begin{align}
    \gammahat^{\text{\tiny ML-NCP}}_n &= \frac{( \ccp(\theta_n, \tauu_n))^\hermit \yyp_n   }{ \norm{ \ccp(\theta_n, \tauu_n)}_2^2 } ~,
\end{align}
leading to the compressed log-likelihood cost function
\begin{align} \label{eq_direct_pos_lik_alt2}
         \llr_{\text{\tiny ML-NCP}}(\pptwoD, \deltatau) &=
         \sum_{n=0}^{N-1} \normbig{ \projnull{\ccp(\theta_n, \tauu_n)} \yyp_n }_2^2 ~,
\end{align}
where $\projnull{\ccp(\theta_n, \tauu_n)} = \bm{I} - \frac{\ccp(\theta_n, \tauu_n) \ccp(\theta_n, \tauu_n)^\mathsf{H}}{\|\ccp(\theta_n, \tauu_n) \|^2}$ is the orthogonal projector onto the null space spanned by $\ccp(\theta_n, \tauu_n)$.
Accordingly, the final expression of the direct ML estimator that does not exploit \ac{CP} information is
\begin{align}\label{eq_direct_pos_final_nocarphase}
         [\hat{\pp}_{\text{\tiny 2D}}^{\text{\tiny ML-NCP}} \ \hat{\delta}_\tau^{\text{\tiny ML-NCP}}] = \arg \min_{\pptwoD,\delta_\tau} ~\llr_{\text{\tiny ML-NCP}}(\pptwoD, \deltatau)   ~.
    \end{align}
Comparing \eqref{eq_direct_pos_lik_alt2} to \eqref{eq_direct_pos_sc1_4}, we immediately notice in \eqref{eq_direct_pos_lik_alt2} the absence of a \textit{cross-RS correlation term} represented by the last term in \eqref{eq_direct_pos_sc1_4}. This term links the different known \acp{RS} positions and the unknown UE position through a common phase offset $\deltaphi$, allowing us to exploit the \ac{CP} information collectively available at all \acp{RS} to infer information about the UE location. To solve \eqref{eq_direct_pos_final_nocarphase}, we follow exactly the same rationale used for \eqref{eq_direct_pos_final}.

\newcommand{\etachtrp}{\etab^{\rm{ch}\trp}}
\newcommand{\efim}{\bm{J}}
\newcommand{\fim}{ {\boldsymbol{J}} }
\newcommand{\fimetab}{\fim_{\etab}}
\newcommand{\fimetach}{\fim_{\etach}}
\newcommand{\fimetachij}{\fim_{\etach_{ij}}}
\newcommand{\fimetachji}{\fim_{\etach_{ji}}}
\newcommand{\fimpos}{\fim_{\pptwoD}}
\newcommand{\fimclock}{\fim_{\mathrm{C}}}
\newcommand{\muyy}{ {\boldsymbol{\mu}_{\yy}} }
\newcommand{\muyyn}{ {\boldsymbol{\mu}_n} }
\newcommand{\aaap}{ \dt{\boldsymbol{\aaa}} }
\newcommand{\bbd}{ \dot{\boldsymbol{\bb}} }
\newcommand{\omegac}{\omega_c} 

\newcommand{\jacobian}{ {\boldsymbol{T}} }
\newcommand{\jacobP}{ {\boldsymbol{P}} }
\newcommand{\jacobC}{ {\boldsymbol{C}} }
\newcommand{\eba}{ {\boldsymbol{e}^\alpha} } 
\newcommand{\range}{ {\boldsymbol{r}} } 

\newcommand{\peb}{\mathcal{P}} 
\newcommand{\ceb}{\mathcal{C}} 
\newcommand{\coeb}{\ceb_{\tau}} 
\newcommand{\cpeb}{\ceb_{\phi}} 

\newcommand{\etabw}{\etab_{\mathrm{w}}}
\newcommand{\etabu}{\etab_{\mathrm{u}}}
\newcommand{\etabclock}{\widetilde{\etab}}
\newcommand{\etabwclock}{\etabclock_{\mathrm{w}}}
\newcommand{\etabuclock}{\etabclock_{\mathrm{u}}}
\newcommand{\deltac}{\bm{\delta}_\mathrm{c}}

\newcommand{\fimA}{ \fim_{\etabw \etabw} }
\newcommand{\fimB}{ \fim_{\etabw \etabu} }
\newcommand{\fimD}{ \fim_{\etabu \etabu} }
\newcommand{\fimAu}{ {\fim}_{\etabwclock \etabwclock} }
\newcommand{\fimBu}{ {\fim}_{\etabuclock \etabwclock} }
\newcommand{\fimDu}{ {\fim}_{\etabuclock \etabuclock} }
\section{Cram\'er-Rao Lower Bound}\label{sec::CRLB}
In this section, we adopt the theoretical tool of the \ac{CRB} to investigate the achievable accuracy in terms of joint positioning and synchronization. 
The \ac{CRB} for the specific problem at hand is defined as \cite{Kay93Estimation}
\begin{align}\label{eq:CRLB_def}
    \mathbb{E}_\etab \left[ (\hat{\etab} - \etab) (\hat{\etab} - \etab)^\trp \right] \succeq \fimetab^{-1}
\end{align}
where $\fimetab$ is the \ac{FIM} for the  vector $\etab = [\pptwoD \ \deltac^\trp \ \alphabar^\trp]^\trp$ containing the \ac{UE} position (in 2D), and nuisance parameters in form of the clock and phase synchronization parameters in $\deltac$ and the \ac{LoS} amplitudes for each \ac{RS}-\ac{UE} link $\alphabar$. 
We assume that each \ac{RS} $n$ contributes independent information on $\etab$, i.e., assuming identical \ac{DMC} and noise statistics, the \ac{FIM} for the joint positioning and synchronization problem is the sum of $N$ contributions   
\begin{align}\label{eq:fim_sum}
    \fimetab = \sum\limits_{n=0}^{N-1} \fimetab^{(n)} = \sum\limits_{n=0}^{N-1} \jacobian_n \fimetach^{(n)}  \jacobian_n^\trp
\end{align}
where $\fimetab^{(n)}$ represents the \ac{FIM} contribution provided by the $n$-th \ac{RS}, which are related to the channel parameter \ac{FIM} $\fimetach^{(n)}$ of each \ac{RS} via the corresponding Jacobian matrix $\jacobian_n$. 
The channel parameter vector of the $n$-th \ac{RS} is defined as 
\begin{equation}
\etach_n = [ \theta_n \ \tauu_n \ \phi_n \ \alpha_n ]^\trp \in \realset{4}{1} \label{eq:fim_etach}
\end{equation}
containing the delay $\tauu_n$, the \ac{AoA} $\theta_n$, the amplitude $\alpha_n$ and phase $\phi_n$ associated to the $n$-th \ac{LoS} path. 
The elements of the  \ac{FIM} $\fimetach^{(n)}$ in~\eqref{eq:fim_sum} are defined as~\cite[Sec.~15.7]{Kay93Estimation}
\begin{align}\label{eq:fim}
    [\fimetach^{(n)}]_{i,j} = 2\realp{
        \frac{\partial \muyyn^\hermit}{\partial [\etach_n]_i} \rrb^{-1} \frac{\partial \muyyn}{\partial [\etach_n]_j}
    }
\end{align}
where $\muyyn = \gamma_n \cc(\theta_n, \tauu_n)$.

To gain insights on the achievable performance, we investigate two different scenarios: i) first, we consider the case of coherent \acp{RS}, which allows us to perform positioning by exploiting the \ac{CP}, and ii) second the case of non-coherent \acp{RS} where \ac{NCP} can be exploited.
Due to \eqref{eq:fim_sum}, both cases can be analyzed by a suitable definition of the synchronization parameters $\deltac$ and consequently the Jacobian matrices $\jacobian_n$
\begin{align}\label{eq:fim-jacobian-compact}
    \jacobian_n \!=\! \frac{\partial \etachtrp_n}{\partial \bm{\eta}} \!=\!\! 
    \begin{bmatrix} 
        \jacobP^{\theta}_n & \jacobP^{\tauu}_n & \jacobP^{\phi}_n & \bm{0}  \\  
        \bm{0} & \bm{C}^{\tauu}_n & \bm{C}^{\phi}_n & \bm{0}  \\ 
        \bm{0} & \bm{0} & \bm{0} & \bm{1}_n  \\  
    \end{bmatrix} \!\in \mathbb{R}^{N+N_\mathrm{c}+2\times 4}.
\end{align}
The block matrices relating channel parameters to the position are found as $\jacobP^{\tauu} = \frac{\range_n}{c \lVert\range_n\rVert}$ from the derivative of~\eqref{eq_taun} w.r.t. position  and as $\jacobP^{\phi} = \frac{-2\pi \, \range_n}{\lambda \, \lVert\range_n\rVert}$ from the derivative of \eqref{eq_phin} (inserting \eqref{eq_taun}) w.r.t. position.
The selection vector $\bm{1}_n$ 
associates the amplitude $\alpha_n$ in $\etach_n$ with the respective amplitude in $\alphabar$ in $\etab$. The definition for $\jacobP^{\theta}$ is given in Appendix~\ref{app_jacobian}.

\paragraph{Coherent / CP} For the coherent case, the synchronization parameter vector contains clock and phase offset parameters $\deltac=[\delta_\tau \ \delta_\phi]^\trp$ which are the same for all \acp{RS}, resulting in $N_\mathrm{c}=2$ synchronization parameters. 
The corresponding block-matrices relating the phase information of the $n$-th \ac{RS} to phase offset and clock offset are found to  be $\jacobC^{\tauu}=[ 1 \ 0 ]^\trp$ and $\jacobC^{\phi}=[ 0 \ 1 ]^\trp$.

\paragraph{Non-coherent / NCP} For the non-coherent case, the \ac{CP} cannot be exploited for positioning, as each \ac{RS} is assumed to have a separate phase offset $\delta_{\phi,n}$ that prevents linking the unknown \ac{UE} position with the known \ac{RS} positions through the phase of the received signal.
The synchronization parameter vector becomes $\deltac=[\delta_\tau \ \bm{\delta}_\phi]^\trp$ with $\bm{\delta}_\phi = [\delta_{\phi,1}\cdots\delta_{\phi,N}]^\trp$, resulting in $N_\mathrm{c}=N+1$ synchronization parameters. 
Consequently, one obtains a block-matrix for each phase offset parameters $\jacobC^{\phi}_n = [0 \ \bm{1}_n^\trp]^\trp$ as the derivatives of \eqref{eq_phin} w.r.t. the clock parameters and $\deltaphi$ are $1$.
Similarly, one obtains $\jacobC^{\tau}_n = [1 \ \bm{0}_{N}^\trp]^\trp$ for the block-matrix of the clock offset.

\subsection{Bounds for Positioning and Synchronization}
To compute the bounds for positioning and synchronization from the \ac{FIM} $\fimetab$ in~\eqref{eq:fim_sum}, we partition the parameter vector as $\etab = [ \etabw^\trp \ \etabu^\trp ]^\trp$, with $\etabw = [\pptwoD \ \deltatau]$ containing the parameters of interest and $\etabu$ all remaining parameters as nuisance parameters~\cite{VanTrees2002optimumASP}.
Block-partitioning of the \ac{FIM}
\begin{align}\label{eq:fim_partitioning}
    \fimetab = 
    \begin{bmatrix}
        \fimA & \fimB \\ \fimB^\trp & \fimD
    \end{bmatrix} \in \realset{(N+N_\mathrm{c}+2)}{(N+N_\mathrm{c}+2)}
\end{align}
allows to make use of the notion of the \ac{EFIM}~\cite{ShenTIT2010,HanTIT2016} to obtain $\efim_{\mathrm{e}} = \fimA - \fimB\fimD^{-1}\fimB^\trp  \in \realset{3}{3}$ \cite{VanTrees2002optimumASP}, 
The \ac{PEB} and \ac{CEB} are then defined as 
\begin{align}
    \peb = \sqrt{ \tracee{[\efim_{\mathrm{e}}^{-1}]_{1:2,1:2}}}, \qquad \coeb = \sqrt{[ \efim_{\mathrm{e}}^{-1} ]_{3,3}}.
\end{align}

\section{Simulation Results}\label{sec::simulations}
In this section, we conduct numerical simulation campaigns to investigate the theoretical performance achievable in terms of both localization and synchronization via inspection of the CRLB expressions derived in Sec.~\ref{sec::CRLB}, as well as to assess the actual estimation performance provided by the novel ML-based estimators designed in Sec.~\ref{sec::design_estimators}. 

\subsection{Scenario}
The considered scenario consists of a network of $N = 4$ \acp{RS} deployed over a squared area of $10 \times 10$ m
placed at the height of \SI{5}{\metre} from the floor level, and of a single \ac{UE} located at $\pp = [7 \ 3 \ 1]^\mathsf{T}$. The \acp{RS} are distributed on each corner of the perimeter so as to provide uniform coverage of the area. Each individual RS is equipped with a ULA with $M = 4$ antennas spaced at $d=\lambda/2$. The \ac{UE} transmits OFDM signals in the uplink channel at a carrier frequency $f_c = \SI{3.5}{\giga\hertz}$, with a bandwidth $B = \SI{100}{\mega\hertz}$, over $K = 100$ different subcarriers. We set the UE clock offset and phase offset to $\delta_{\tau} = 100/c\,\SI{}{\second}$
and $\delta_\phi = \SI{10}{\SIdeg}$, respectively. The channel amplitudes are generated as $\alpha_n = \sqrt{P} \rho_n$ where $P$ denotes the \ac{UE} transmit power and $\rho_n$ is set according to the common path loss model in free space, i.e., $\rho_n = \lambda/(4\pi \|\pp - \pprs_n\|)$, whereas the noise power is $\sigma^2 = k_BT_0 B$, $k_B$ being the Boltzmann constant and $T_0$ the standard thermal noise temperature. As to the parameters of the \ac{DMC}, we set the normalized coherence bandwidth $\beta_d = 1/(T_d B)$ with $T_d = 20/c$ decay time set to a distance of \SI{20}{\metre}, the normalized onset time to 
$\tau_d = \frac{B}{K}\left(\tau_n + \SI{1}{\metre}/c\right)$,
i.e., the \ac{DMC} onset is delayed by \SI{1}{\metre} w.r.t. the \ac{LoS},
and the peak power $\alpha_d$ is chosen to guarantee a \ac{DNR}, defined as $\mathrm{DNR} = \alpha_d/ \sigma^2$, equal to \SI{20}{\dB}. To quantify the average signal power received by the whole \ac{RS} network, we define an average \ac{SDNR} as\footnote{This definition should be seen as a positioning-specific metric, being the additional \ac{DMC} contributions treated as disturbance terms. Conversely, for communication-oriented tasks, \ac{DMC} can be included in the useful signal part (i.e., summing up with the signal power at the numerator of \eqref{def:SDNR}) since they can be constructively exploited to convey additional information.}
\begin{equation}\label{def:SDNR}
\overline{\mathrm{SDNR}} =  \frac{P}{N K}\sum_{n=0}^{N-1}\rho^2_n \  \cc^\hermit \rrb_n^{-1}\cc
\end{equation}  
where index $n$ is added to $\rrb$ to make explicit that also the \ac{DMC} power will vary between \acp{RS}, in realistic scenarios.

\subsection{Analysis of the Positioning Bounds}
\begin{figure}[t]
    \centering\hspace{-0.5mm}
    \setlength{\plotWidth}{0.85\linewidth}
    \setlength{\plotHeight}{0.3\linewidth}
    \input{Figures/bounds-sweep.tex}\vspace{-6mm}
    \caption{\ac{PEB} as a function of the bandwidth $B$ for different number of antennas $M$. The comparison includes \acp{RS} either exploiting carrier-phase (CP) information or ignoring
    it (NCP), at a fixed $\overline{\mathrm{SDNR}}\approx\SI{12}{\dB}$.}
     \label{fig:bounds-sweep}\vspace{-4mm}
\end{figure}
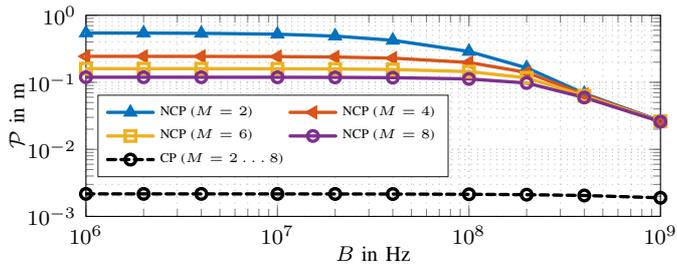

We start by analyzing the \acp{PEB} derived in Sec.~\ref{sec::CRLB}  to investigate how the number of antennas $M$ per \ac{RS} and bandwidth $B$ impact on the ultimate positioning accuracy. 
Fig.~\ref{fig:bounds-sweep} shows these bounds evaluated for both positioning with carrier-phase (CP) and without (NCP).
During the analysis, we keep a constant $\overline{\mathrm{SDNR}} \approx \SI{12}{\dB}$, which cancels the effect of a varying array gain when varying $M$. The first important fact can be highlighted by comparing the curves corresponding to the two groups of \acp{PEB}: the evident gap between the case in which \ac{CP} information is exploited, and the case where it is not, clearly demonstrates the crucial role that such information has on the positioning accuracy. As a matter of fact, exploiting \ac{CP} information brings about two orders of magnitude improvements in the ultimate UE localization accuracy, with errors that can be as low as a few millimeters. Delving into more specific details, in the NCP case we observe that the array aperture, i.e., its angular resolution, dominates the \ac{PEB} for low bandwidths.
For high bandwidths, the \ac{PEB} is instead dominated by its high time resolution.
In the CP case, interestingly, varying the number of antennas or bandwidth has a negligible impact on the \ac{PEB}. 
This behavior is linked to the fact that the accuracy from phase-aided positioning exploiting the fully coherent \ac{RS} infrastructure dominates over both its time and angular resolutions. 
 The likelihood function underlying the \ac{CP}-based positioning  exhibits very sharp peaks at the spatial intersections of the wavefronts associated to each RS.
 The very informative peak around the true position $\pptwoD$, however, comes at the price of a multimodal likelihood function, and thus a computationally more demanding estimation problem. More quantitatively, the resolution of \eqref{eq_direct_pos_final} involves a computational cost about $38\%$ higher than the cost needed to solve \eqref{eq_direct_pos_final_nocarphase}.

\subsection{Algorithms Performance Assessment}
\begin{figure}[t]
     \setlength{\plotWidth}{0.85\linewidth}
    \setlength{\plotHeight}{0.37\linewidth}
    \input{Figures/RMSE_PEB.tex}\vspace{-6mm}
     \caption{RMSE on UE position estimation compared to PEB as a function of the $\overline{\mathrm{SDNR}}$, either exploiting \ac{CP} or ignoring it (\ac{NCP}).}
     \label{fig:RMSE_PEB}\vspace{-2mm}
     \end{figure}
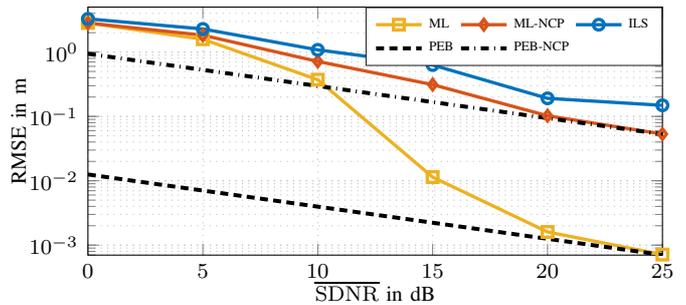

We now assess the performance of the ML-based estimation algorithms developed in Sec.~\ref{sec::design_estimators}. In Fig.~\ref{fig:RMSE_PEB}, we report the \ac{RMSE} on the estimation of the \ac{UE} position $\pptwoD$, obtained by averaging the results over \num{1000} independent Monte Carlo trials, as a function of the $\overline{\mathrm{SDNR}}$. The comparison includes the ML algorithm that exploits the \ac{CP}  information, its relaxed version ML-NCP that ignores the existence of a relationship between the phase of the signal received at each \ac{RS} and the unknown \ac{UE} position, and the theoretical lower bounds derived in Sec.~\ref{sec::CRLB} acting as \emph{benchmark}. For completeness, we also report the performance
of the \ac{ILS} estimator which is used to obtain an initial estimate of both UE position and synchronization offset. 

 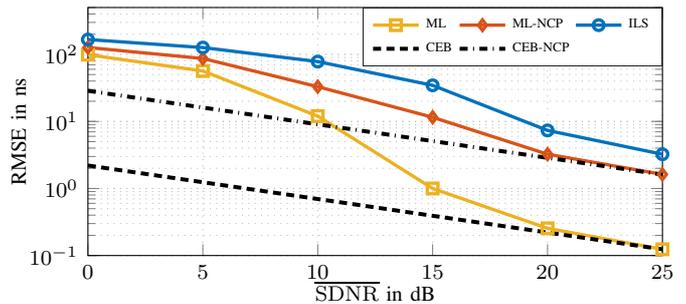
\begin{figure}[t]
    \setlength{\plotWidth}{0.85\linewidth}
    \setlength{\plotHeight}{0.37\linewidth}
    \input{Figures/RMSE_CEB.tex}\vspace{-6mm}
    \caption{RMSE on clock-offset estimation compared to CEB as a function of the $\overline{\mathrm{SDNR}}$, either exploiting \ac{CP} or ignoring it (\ac{NCP}).}
     \label{fig:RMSE_CEB}\vspace{-4mm}
 \end{figure}

From an algorithmic perspective, it can be noticed that all the approaches start by exhibiting quite high \acp{RMSE} for low values of the $\overline{\mathrm{SDNR}}$. This behavior can be explained by observing that the suboptimal \ac{ILS} approach that is used to initialize the ML estimators is not very accurate in such a regime. However, as soon as the initialization provided by  the \ac{ILS} improves, the \acp{RMSE} of the ML estimators immediately drop and tend to approach the corresponding lower bounds. Remarkably, the proposed ML estimator that exploits \ac{CP} information achieves an accuracy in the order of about \SI{1}{\centi\metre} for $\overline{\mathrm{SDNR}} = \SI{15}{\dB}$, and further enhances until mm-level accuracy as the $\overline{\mathrm{SDNR}}$ increases. A similar trend is observed in the \ac{RMSE} of the ML-NCP, which however requires a $\overline{\mathrm{SDNR}}$ of about \SI{20}{\dB} to attain values in the order of \SI{10}{\centi\metre}.

 To complement the above analysis, in Fig.~\ref{fig:RMSE_CEB} we evaluate the performance on the estimation of the \ac{UE} clock offset $\delta_\tau$ (similar results are obtained also for the phase offset $\delta_\phi$, hence they are omitted). A direct comparison of the CEBs confirms that the use of \ac{CP} information has a beneficial effect also on the synchronization accuracy, with improvements of about one order of magnitude compared to the case in which such  information is not exploited. Consistently with the results in Fig.~\ref{fig:RMSE_PEB}, also in this case the algorithms exhibit higher \ac{RMSE} values in the lower $\overline{\mathrm{SDNR}}$ regime, due to the poor accuracy of the initialization provided by the \ac{ILS} algorithm. Interestingly, the proposed ML estimator that leverages \ac{CP} information guarantees an accurate synchronization in the order of \SI{1}{\nano\second} already for $\overline{\mathrm{SDNR}} \geq \SI{15}{\dB}$, significantly outperforming the ML-NCP algorithm that instead needs about \SI{10}{\dB} more to achieve a comparable level of accuracy.

\section{Conclusion}
We addressed the problem of  high accuracy positioning exploiting coherent processing performed by distributed \acp{RS}, seen as a promising technology for future wireless networks.
To take full advantage of the large aperture achieved by the  spatial distribution of the \acp{RS}, accurate time- and phase- synchronization need to be recovered.
To this aim, we proposed a novel ML algorithm that jointly estimates the synchronization parameters alongside the \ac{UE} position, for a propagation scenario consisting of an \ac{LoS}-path that can be exploited for positioning and \ac{DMC} modeling the disturbance due to diffuse multipath. 
Two variants of the ML estimator have been derived, assuming either a fully coherent or a non-coherent \ac{RS} network.
The performance of both variants is compared to the corresponding \acp{CRB}.
The obtained results show that the coherent ML algorithm significantly outperforms its non-coherent counterpart.
Moreover, exploiting the \ac{CP} allows to perform high accuracy positioning even with fewer per-\ac{RS} antenna elements, as the position-related information conveyed from the \ac{CP} is much richer and compensates for the reduced aperture gain.

\begin{appendices}

\section{Jacobian Matrix: Block Matrix for \ac{AoA}}\label{app_jacobian}
To obtain $\jacobP^\theta$ we denote the vector from the $n$-th \ac{RS} to the \ac{UE} as $\range_n = [r_{xn} \ r_{yn} ]^\trp$ 
with $r_{xn} = p_x - x_n^\mathrm{RS}$ and $r_{yn} = p_y - y_n^\mathrm{RS}$.
For \acp{RS} horizontally non-colocated with the \ac{UE}, i.e., $\range_n \neq \bm{0}$, 
one obtains
\begin{align}\label{eq_jacobP}
    \jacobP^\theta \!=\!
    \begin{cases} 
        \begin{bmatrix}
            \frac{r_{yn}}{r_{xn}^2+r_{yn}^2} & \frac{-r_{xn}}{r_{xn}^2+r_{yn}^2} 
        \end{bmatrix}^\trp , & [\pps_{n}]_1 \neq 0
        \\
        \bm{0} \, , & [\pps_{n}]_1 = 0, [\pps_{n}]_2 \neq 0
    \end{cases}
\end{align}
through the position-related derivatives of~\eqref{eq_thetanl}.

\section{Fisher Information Matrix}\label{app_fim}
The elements of the \ac{FIM} $\fimetach$ in~\eqref{eq:fim} are given below, omitting the index $n$ for brevity.
Due to symmetry, $[\fimetach]_{j,i} = [\fimetach]_{i,j}$ holds and the \ac{FIM} is described through its elements 
\begin{align*}
     \fim_{\theta,\theta} &= 2\realp{
        \alpha_n^{2} \dot{\cc}_{\theta}^\hermit \rrb^{-1} \dot{\cc}_{\theta}
    }  &
    \fim_{\theta,\tauu} &= 0  \\ 
    \fim_{\theta,\phi} &= 0  & 
    \fim_{\theta,\alpha} &= 0 \\
    \fim_{\tauu,\tauu} &= 2 \Re 
        \left\{
            \alpha_n^2 \dot{\cc}_{\tauu}^\hermit \rrb^{-1} \dot{\cc}_{\tauu}
        \right\}  &
    \fim_{\tauu,\phi} &= 2 \Re \left\{
          j \alpha_n^2 \dot{\cc}_{\tauu}^\hermit \rrb^{-1} \cc
    \right\} \\
    \fim_{\tauu,\alpha} &= 2 \Re \left\{
         \alpha_n \dot{\cc}_{\tauu}^\hermit \rrb^{-1} \cc
    \right\}   &
    \fim_{\phi,\phi} &= 2\realp{\alpha_n^2 \cc^\hermit \rrb^{-1} \cc} \\
    \fim_{\phi,\alpha} &= 0
    &
    \fim_{\alpha,\alpha} &= 2\realp{\cc^\hermit \rrb^{-1}\cc} 
\end{align*}with $\cc=(\bb \odot \sss) \otimes \aaa$, $\dot{\cc}_{\theta} = \frac{\partial \cc}{\partial \theta}= (\bb \odot \sss) \otimes \aaap$ and $\dot{\cc}_{\tauu} = \frac{\partial \cc}{\partial \tauu}=(\bbd \odot \sss) \otimes \aaa$.
Notice that due to the selection of the array reference point, the elements $\fim_{\theta,\tauu}$, $\fim_{\theta,\phi}$ and $\fim_{\theta,\alpha}$ vanish.

\end{appendices}

\section*{Acknowledgments}
{\color{black}This work was supported, in part, by the Swedish Research Council project 2022-03007, and in part by
the European Union’s Horizon 2020 research and innovation programme under grant agreement No 101013425 (Project ``REINDEER").} 

\bibliographystyle{IEEEtran}
\balance
\bibliography{ICC_Workshop}

\end{document}

%% file: Other/commands.tex
\theoremstyle{remark}

\newtheoremstyle{mytheoremstyle} 
    {\topsep}                    	
    {\topsep}                    	
    {\upshape}                 	
    {.5em}                        	
    {\itshape}                   	
    {.}                          		
    {.5em}                       	
    {}  

\theoremstyle{plain}

\newtheoremstyle{iremark}
  {\topsep}   
  {\topsep}   
  {\upshape}  
  {0.2in}       
  {\itshape}  
  {.}         
  {5pt plus 1pt minus 1pt} 
  {\thmname{#1}\thmnumber{ \itshape#2}\thmnote{ (#3)}} 

\newtheorem{proposition}{Proposition}

\theoremstyle{definition}
\newtheorem*{proof}{Proof}

\DeclareFontFamily{U}{mathx}{\hyphenchar\font45}
\DeclareFontShape{U}{mathx}{m}{n}{
	<5> <6> <7> <8> <9> <10>
	<10.95> <12> <14.4> <17.28> <20.74> <24.88>
	mathx10
}{}
\DeclareSymbolFont{mathx}{U}{mathx}{m}{n}
\DeclareFontSubstitution{U}{mathx}{m}{n}

\DeclarePairedDelimiter\absbigg{\Bigg\lvert}{\Bigg\rvert}%

\makeatletter \renewcommand\d[1]{\ensuremath{%
		\;\mathrm{d}#1\@ifnextchar\d{\!}{}}}
\makeatother

\makeatletter
\newcommand*\rel@kern[1]{\kern#1\dimexpr\macc@kerna}
\newcommand*\widebar[1]{%
  \begingroup
  \def\mathaccent##1##2{%
    \rel@kern{0.8}%
    \overline{\rel@kern{-0.8}\macc@nucleus\rel@kern{0.2}}%
    \rel@kern{-0.2}%
  }%
  \macc@depth\@ne
  \let\math@bgroup\@empty \let\math@egroup\macc@set@skewchar
  \mathsurround\z@ \frozen@everymath{\mathgroup\macc@group\relax}%
  \macc@set@skewchar\relax
  \let\mathaccentV\macc@nested@a
  \macc@nested@a\relax111{#1}%
  \endgroup
}
\makeatother

\usepackage{scalerel,stackengine}
\stackMath
\newcommand\widecheck[1]{%
\savestack{\tmpbox}{\stretchto{%
  \scaleto{%
    \scalerel*[\widthof{\ensuremath{#1}}]{\kern-.6pt\bigwedge\kern-.6pt}%
    {\rule[-\textheight/2]{1ex}{\textheight}}
  }{\textheight}%
}{0.5ex}}%
\stackon[1pt]{#1}{\scalebox{-1}{\tmpbox}}%
}

\newcommand{\DMC}{\text{\tiny DMC}}
\newcommand{\RS}{\text{\tiny RS}}

\newcommand{\projnull}[1]{\boldsymbol{\Pi}^{\perp}_{#1}}

\newcommand{\norm}[1]{\left\lVert#1\right\rVert}

\newcommand{\normbig}[1]{\Big\lVert#1\Big\rVert}

\newcommand{\pp}{\bm{p}}
\newcommand{\pptwoD}{\bm{p}_{\text{\tiny 2D}}}
\newcommand{\hh}{\bm{h}}
\newcommand{\ww}{\bm{w}}
\newcommand{\aaa}{\bm{a}}
\newcommand{\bb}{\bm{b}}
\newcommand{\cc}{\bm{c}}

\newcommand{\yy}{\bm{y}}
\newcommand{\sss}{\bm{s}}
\newcommand{\zz}{\bm{z}}
\newcommand{\xx}{\bm{x}}

\newcommand{\pps}{\pp^{\prime}}
\newcommand{\yyp}{\yy^{\prime}}

\newcommand{\ccp}{\cc^{\prime}}

\newcommand{\yybr}{\Breve{\yy}}
\newcommand{\ccbr}{\Breve{\cc}}

\newcommand{\llr}{\mathcal{L}}

\newcommand{\yyb}{\bm{Y}}
\newcommand{\wwb}{\bm{W}}
\newcommand{\zzb}{\bm{Z}}
\newcommand{\rrb}{\bm{R}}

\newcommand{\mmb}{\bm{M}}

\newcommand{\deltaf}{\Delta_f}
\newcommand{\deltaphi}{\delta_{\phi}}
\newcommand{\deltatau}{\delta_{\tau}}
\newcommand{\deltaphihat}{\widehat{\delta}_{\phi}}
\newcommand{\tauu}{\widetilde{\tau}}

\newcommand{\pprs}{\pp^{\RS}}
\newcommand{\xrs}{x^{\RS}}
\newcommand{\yrs}{y^{\RS}}
\newcommand{\zrs}{z^{\RS}}

\newcommand{\wwdmc}{\ww^{\DMC}}
\newcommand{\wwbdmc}{\wwb^{\DMC}}
\newcommand{\rrbdmc}{\rrb^{\DMC}}

\newcommand{\toep}{{\rm{Toep}}}

\newcommand{\trp}{\mathsf{T}}
\newcommand{\hermit}{\mathsf{H}}
\newcommand{\fc}{f_c}
\newcommand{\mtCN}{{\mathcal{CN}}}

\newcommand{\etadmc}{\etab_{\DMC}}
\newcommand{\etach}{\etab^{\rm{ch}}}

\newcommand{\complexset}[2]{ \mathbb{C}^{#1 \times #2}  }

\newcommand{\complexsett}{ \mathbb{C}  }
\newcommand{\realset}[2]{ \mathbb{R}^{#1 \times #2}  }
\newcommand{\realsetone}[1]{ \mathbb{R}^{#1}  }

\newcommand{\intset}{ \mathbb{Z}  }

\newcommand{\Imatrix}{{ \bm{\mathrm{I}} }}
\newcommand{\boldzero}{{ {\bm{0}} }}
\newcommand{\boldone}{{ {\bm{1}} }}

\newcommand{\realp}[1]{ \Re \left\{#1\right\}  }
\newcommand{\imp}[1]{ \Im \left\{#1\right\}  }
\newcommand{\vecc}[1]{ {\rm{vec}}\left(#1\right)  }

\newcommand{\etab}{ {\boldsymbol{\eta}} }
\newcommand{\etabhat}{ \widehat{\etab} }
\newcommand{\kappab}{ {\boldsymbol{\kappa}} }
\newcommand{\alphab}{ {\boldsymbol{\alpha}} }
\newcommand{\gammab}{ {\boldsymbol{\gamma}} }
\newcommand{\alphahat}{ {\widehat{\alpha}} }

\newcommand{\alphabar}{ \widebar{\alphab} }

\newcommand{\veccinv}[1]{ {\rm{reshape}}_{M,K}\left(#1\right)  }

\newcommand{\gammahat}{\widehat{\gamma}}

\newcommand{\tracee}[1]{ {{ {\rm{tr}} \big\{ #1 \big\} }}  }

%% file: Figures/scenario.tex
%
%

\pgfplotsset{every axis/.append style={
  label style={font=\footnotesize},
  legend style={font=\footnotesize},
  tick label style={font=\footnotesize},
  xticklabel={
    \ifdim \tick pt < 0pt
      \pgfmathparse{abs(\tick)}%
      \llap{$-{}$}\pgfmathprintnumber{\pgfmathresult}
   \else
      \pgfmathprintnumber{\tick}
   \fi
}}}

\definecolor{mycolor1}{rgb}{0.68630,0.19610,0.20780}%
\definecolor{mycolor2}{rgb}{0.66270,0.20390,0.12160}%
\definecolor{mycolor3}{rgb}{0.00000,0.40390,0.58430}%
\definecolor{mygray}{rgb}{0.2,0.2,0.2}%
\begin{tikzpicture}

\begin{axis}[%
width=0.805\plotWidth,
height=0.5\plotWidth,
at={(0\plotWidth,0\plotWidth)},
scale only axis,
tick align=inside,		       
axis line style = thick,	   
line cap = round,
plot box ratio=1.13 1.043 1,
colormap/blackwhite,
xmin=0,
xmax=2.6,
xlabel={$x$ in \SI{}{\metre}},
ymin=0,
ymax=2.4,
ylabel={$y$ in \SI{}{\metre}},
zmin=0,
zmax=2.3,
zlabel={$z$ in \SI{}{\metre}},
view={110}{20},
axis background/.style={fill=white},
ytick={0.5,1,...,2},                  
xmajorgrids,
ymajorgrids,
zmajorgrids,
grid style=dotted
]

\addplot3[%
surf,
fill opacity=0.3, fill=white!90!black, shader=faceted,   
color=white!20!black, z buffer=sort, colormap/blackwhite, mesh/rows=5]
table[row sep=crcr, point meta=\thisrow{c}] {%
x	y	z	c\\
2.77555756156289e-17	0.75	0	0\\
2.77555756156289e-17	0.75	0.35	0.35\\
0.35	0.75	0	0\\
0.35	0.75	0.35	0.35\\
0.35	2.25	0	0\\
0.35	2.25	0.35	0.35\\
2.77555756156289e-17	2.25	0	0\\
2.77555756156289e-17	2.25	0.35	0.35\\
-2.77555756156289e-17	0.75	0	0\\
-2.77555756156289e-17	0.75	0.35	0.35\\
};

\addplot3[area legend, table/row sep=crcr, patch, patch type=polygon, vertex count=5, fill opacity=0.3, 
fill=white!90!black, faceted color=white!20!black, forget plot, patch table with point meta={%
0	1	2	3	4	1\\
5	6	7	8	9	1\\
}]
table[row sep=crcr] {%
x	y	z\\
2.77555756156289e-17	0.75	0.35\\
0.35	0.75	0.35\\
0.35	2.25	0.35\\
2.77555756156289e-17	2.25	0.35\\
-2.77555756156289e-17	0.75	0.35\\
2.77555756156289e-17	0.75	0\\
0.35	0.75	0\\
0.35	2.25	0\\
2.77555756156289e-17	2.25	0\\
-2.77555756156289e-17	0.75	0\\
};

\addplot3[area legend, table/row sep=crcr, patch, patch type=polygon, vertex count=5, fill opacity=0.3, fill=white!20!black, faceted color=white!20!black, forget plot, patch table with point meta={%
0	1	2	3	4	1\\
5	6	7	8	9	1\\
}]
table[row sep=crcr] {%
x	y	z\\
2.77555756156289e-17	0.75	0\\
0.35	0.75	0\\
0.35	2.25	0\\
2.77555756156289e-17	2.25	0\\
-2.77555756156289e-17	0.75	0\\
2.77555756156289e-17	0.75	0\\
0.35	0.75	0\\
0.35	2.25	0\\
2.77555756156289e-17	2.25	0\\
-2.77555756156289e-17	0.75	0\\
};
\addplot3 [color=mycolor2, loosely dashed, line cap = round, line width=0.7pt]
 table[row sep=crcr] {%
1	1.2	1\\
0.35	1.5	0.35\\
};

\addplot3[%
surf, shader=interp,   
fill opacity=0.45, fill=white!90!black, z buffer=sort, colormap/blackwhite, mesh/rows=101]
table[row sep=crcr, point meta=\thisrow{c}] {%
x	y	z	c\\
1.7	0.25	0	0\\
1.7	0.25	0.4	0\\
1.69960534568565	0.262558103905863	0	0\\
1.69960534568565	0.262558103905863	0.4	0\\
1.6984229402629	0.275066646712861	0	0\\
1.6984229402629	0.275066646712861	0.4	0\\
1.69645745014574	0.287476262917145	0	0\\
1.69645745014574	0.287476262917145	0.4	0\\
1.69371663222573	0.299737977432971	0	0\\
1.69371663222573	0.299737977432971	0.4	0\\
1.69021130325903	0.311803398874989	0	0\\
1.69021130325903	0.311803398874989	0.4	0\\
1.68595529717765	0.323624910536936	0	0\\
1.68595529717765	0.323624910536936	0.4	0\\
1.6809654104932	0.335155858313015	0	0\\
1.6809654104932	0.335155858313015	0.4	0\\
1.67526133600877	0.346350734820343	0	0\\
1.67526133600877	0.346350734820343	0.4	0\\
1.6688655851004	0.357165358995799	0	0\\
1.6688655851004	0.357165358995799	0.4	0\\
1.66180339887499	0.367557050458495	0	0\\
1.66180339887499	0.367557050458495	0.4	0\\
1.65410264855516	0.377484797949738	0	0\\
1.65410264855516	0.377484797949738	0.4	0\\
1.64579372548428	0.386909421185738	0	0\\
1.64579372548428	0.386909421185738	0.4	0\\
1.63690942118574	0.395793725484282	0	0\\
1.63690942118574	0.395793725484282	0.4	0\\
1.62748479794974	0.404102648555158	0	0\\
1.62748479794974	0.404102648555158	0.4	0\\
1.61755705045849	0.41180339887499	0	0\\
1.61755705045849	0.41180339887499	0.4	0\\
1.6071653589958	0.418865585100403	0	0\\
1.6071653589958	0.418865585100403	0.4	0\\
1.59635073482034	0.425261336008773	0	0\\
1.59635073482034	0.425261336008773	0.4	0\\
1.58515585831301	0.430965410493204	0	0\\
1.58515585831301	0.430965410493204	0.4	0\\
1.57362491053694	0.43595529717765	0	0\\
1.57362491053694	0.43595529717765	0.4	0\\
1.56180339887499	0.440211303259031	0	0\\
1.56180339887499	0.440211303259031	0.4	0\\
1.54973797743297	0.443716632225726	0	0\\
1.54973797743297	0.443716632225726	0.4	0\\
1.53747626291715	0.446457450145738	0	0\\
1.53747626291715	0.446457450145738	0.4	0\\
1.52506664671286	0.448422940262896	0	0\\
1.52506664671286	0.448422940262896	0.4	0\\
1.51255810390586	0.449605345685654	0	0\\
1.51255810390586	0.449605345685654	0.4	0\\
1.5	0.45	0	0\\
1.5	0.45	0.4	0\\
1.48744189609414	0.449605345685654	0	0\\
1.48744189609414	0.449605345685654	0.4	0\\
1.47493335328714	0.448422940262896	0	0\\
1.47493335328714	0.448422940262896	0.4	0\\
1.46252373708285	0.446457450145738	0	0\\
1.46252373708285	0.446457450145738	0.4	0\\
1.45026202256703	0.443716632225726	0	0\\
1.45026202256703	0.443716632225726	0.4	0\\
1.43819660112501	0.440211303259031	0	0\\
1.43819660112501	0.440211303259031	0.4	0\\
1.42637508946306	0.43595529717765	0	0\\
1.42637508946306	0.43595529717765	0.4	0\\
1.41484414168699	0.430965410493204	0	0\\
1.41484414168699	0.430965410493204	0.4	0\\
1.40364926517966	0.425261336008773	0	0\\
1.40364926517966	0.425261336008773	0.4	0\\
1.3928346410042	0.418865585100403	0	0\\
1.3928346410042	0.418865585100403	0.4	0\\
1.38244294954151	0.41180339887499	0	0\\
1.38244294954151	0.41180339887499	0.4	0\\
1.37251520205026	0.404102648555158	0	0\\
1.37251520205026	0.404102648555158	0.4	0\\
1.36309057881426	0.395793725484282	0	0\\
1.36309057881426	0.395793725484282	0.4	0\\
1.35420627451572	0.386909421185738	0	0\\
1.35420627451572	0.386909421185738	0.4	0\\
1.34589735144484	0.377484797949738	0	0\\
1.34589735144484	0.377484797949738	0.4	0\\
1.33819660112501	0.367557050458495	0	0\\
1.33819660112501	0.367557050458495	0.4	0\\
1.3311344148996	0.357165358995799	0	0\\
1.3311344148996	0.357165358995799	0.4	0\\
1.32473866399123	0.346350734820343	0	0\\
1.32473866399123	0.346350734820343	0.4	0\\
1.3190345895068	0.335155858313015	0	0\\
1.3190345895068	0.335155858313015	0.4	0\\
1.31404470282235	0.323624910536936	0	0\\
1.31404470282235	0.323624910536936	0.4	0\\
1.30978869674097	0.31180339887499	0	0\\
1.30978869674097	0.31180339887499	0.4	0\\
1.30628336777427	0.299737977432971	0	0\\
1.30628336777427	0.299737977432971	0.4	0\\
1.30354254985426	0.287476262917145	0	0\\
1.30354254985426	0.287476262917145	0.4	0\\
1.3015770597371	0.275066646712861	0	0\\
1.3015770597371	0.275066646712861	0.4	0\\
1.30039465431435	0.262558103905863	0	0\\
1.30039465431435	0.262558103905863	0.4	0\\
1.3	0.25	0	0\\
1.3	0.25	0.4	0\\
1.30039465431435	0.237441896094137	0	0\\
1.30039465431435	0.237441896094137	0.4	0\\
1.3015770597371	0.224933353287139	0	0\\
1.3015770597371	0.224933353287139	0.4	0\\
1.30354254985426	0.212523737082855	0	0\\
1.30354254985426	0.212523737082855	0.4	0\\
1.30628336777427	0.200262022567029	0	0\\
1.30628336777427	0.200262022567029	0.4	0\\
1.30978869674097	0.18819660112501	0	0\\
1.30978869674097	0.18819660112501	0.4	0\\
1.31404470282235	0.176375089463064	0	0\\
1.31404470282235	0.176375089463064	0.4	0\\
1.3190345895068	0.164844141686986	0	0\\
1.3190345895068	0.164844141686986	0.4	0\\
1.32473866399123	0.153649265179657	0	0\\
1.32473866399123	0.153649265179657	0.4	0\\
1.3311344148996	0.142834641004201	0	0\\
1.3311344148996	0.142834641004201	0.4	0\\
1.33819660112501	0.132442949541505	0	0\\
1.33819660112501	0.132442949541505	0.4	0\\
1.34589735144484	0.122515202050262	0	0\\
1.34589735144484	0.122515202050262	0.4	0\\
1.35420627451572	0.113090578814262	0	0\\
1.35420627451572	0.113090578814262	0.4	0\\
1.36309057881426	0.104206274515718	0	0\\
1.36309057881426	0.104206274515718	0.4	0\\
1.37251520205026	0.0958973514448421	0	0\\
1.37251520205026	0.0958973514448421	0.4	0\\
1.38244294954151	0.0881966011250105	0	0\\
1.38244294954151	0.0881966011250105	0.4	0\\
1.3928346410042	0.0811344148995969	0	0\\
1.3928346410042	0.0811344148995969	0.4	0\\
1.40364926517966	0.0747386639912273	0	0\\
1.40364926517966	0.0747386639912273	0.4	0\\
1.41484414168699	0.069034589506796	0	0\\
1.41484414168699	0.069034589506796	0.4	0\\
1.42637508946306	0.0640447028223498	0	0\\
1.42637508946306	0.0640447028223498	0.4	0\\
1.43819660112501	0.0597886967409693	0	0\\
1.43819660112501	0.0597886967409693	0.4	0\\
1.45026202256703	0.0562833677742738	0	0\\
1.45026202256703	0.0562833677742738	0.4	0\\
1.46252373708285	0.0535425498542622	0	0\\
1.46252373708285	0.0535425498542622	0.4	0\\
1.47493335328714	0.0515770597371044	0	0\\
1.47493335328714	0.0515770597371044	0.4	0\\
1.48744189609414	0.0503946543143457	0	0\\
1.48744189609414	0.0503946543143457	0.4	0\\
1.5	0.05	0	0\\
1.5	0.05	0.4	0\\
1.51255810390586	0.0503946543143457	0	0\\
1.51255810390586	0.0503946543143457	0.4	0\\
1.52506664671286	0.0515770597371044	0	0\\
1.52506664671286	0.0515770597371044	0.4	0\\
1.53747626291714	0.0535425498542622	0	0\\
1.53747626291714	0.0535425498542622	0.4	0\\
1.54973797743297	0.0562833677742738	0	0\\
1.54973797743297	0.0562833677742738	0.4	0\\
1.56180339887499	0.0597886967409693	0	0\\
1.56180339887499	0.0597886967409693	0.4	0\\
1.57362491053694	0.0640447028223498	0	0\\
1.57362491053694	0.0640447028223498	0.4	0\\
1.58515585831301	0.069034589506796	0	0\\
1.58515585831301	0.069034589506796	0.4	0\\
1.59635073482034	0.0747386639912272	0	0\\
1.59635073482034	0.0747386639912272	0.4	0\\
1.6071653589958	0.0811344148995969	0	0\\
1.6071653589958	0.0811344148995969	0.4	0\\
1.61755705045849	0.0881966011250105	0	0\\
1.61755705045849	0.0881966011250105	0.4	0\\
1.62748479794974	0.0958973514448421	0	0\\
1.62748479794974	0.0958973514448421	0.4	0\\
1.63690942118574	0.104206274515718	0	0\\
1.63690942118574	0.104206274515718	0.4	0\\
1.64579372548428	0.113090578814262	0	0\\
1.64579372548428	0.113090578814262	0.4	0\\
1.65410264855516	0.122515202050262	0	0\\
1.65410264855516	0.122515202050262	0.4	0\\
1.66180339887499	0.132442949541505	0	0\\
1.66180339887499	0.132442949541505	0.4	0\\
1.6688655851004	0.142834641004201	0	0\\
1.6688655851004	0.142834641004201	0.4	0\\
1.67526133600877	0.153649265179657	0	0\\
1.67526133600877	0.153649265179657	0.4	0\\
1.6809654104932	0.164844141686986	0	0\\
1.6809654104932	0.164844141686986	0.4	0\\
1.68595529717765	0.176375089463064	0	0\\
1.68595529717765	0.176375089463064	0.4	0\\
1.69021130325903	0.18819660112501	0	0\\
1.69021130325903	0.18819660112501	0.4	0\\
1.69371663222573	0.200262022567029	0	0\\
1.69371663222573	0.200262022567029	0.4	0\\
1.69645745014574	0.212523737082855	0	0\\
1.69645745014574	0.212523737082855	0.4	0\\
1.6984229402629	0.224933353287139	0	0\\
1.6984229402629	0.224933353287139	0.4	0\\
1.69960534568565	0.237441896094137	0	0\\
1.69960534568565	0.237441896094137	0.4	0\\
1.7	0.25	0	0\\
1.7	0.25	0.4	0\\
};

\addplot3[area legend, draw=white!20!black, fill=white!90!black, fill opacity=0.5,line width=0.5pt, forget plot]
table[row sep=crcr] {%
x	y	z\\
1.7	0.25	0.4\\
1.69959733529438	0.262684783931313	0.4\\
1.69839096256616	0.27531849071475	0.4\\
1.69638573945254	0.287850248872082	0.4\\
1.69358974027927	0.300229597436216	0.4\\
1.69001422354819	0.312406689139697	0.4\\
1.68567358660321	0.324332491132066	0.4\\
1.68058530765732	0.335958982417834	0.4\\
1.67476987541396	0.347239347220094	0.4\\
1.66825070656624	0.35812816349112	0.4\\
1.66105405150621	0.368581585810928	0.4\\
1.6532088886238	0.378557521937308	0.4\\
1.64474680762101	0.388015802296422	0.4\\
1.63570188231143	0.396918341731507	0.4\\
1.6261105334169	0.405229292858351	0.4\\
1.61601138191424	0.412915190410067	0.4\\
1.6054450935221	0.419945085989903	0.4\\
1.59445421495454	0.426290672689516	0.4\\
1.58308300260038	0.431926399070904	0.4\\
1.57137724431837	0.436829572053021	0.4\\
1.55938407506565	0.440980448288815	0.4\\
1.54715178710189	0.444362313664708	0.4\\
1.53472963553339	0.446961550602442	0.4\\
1.5221676399802	0.448767692892251	0.4\\
1.50951638316475	0.449773467836602	0.4\\
1.49682680723304	0.449974825534775	0.4\\
1.48415000862864	0.449370955190388	0.4\\
1.47153703234534	0.447964288376187	0.4\\
1.45903866638696	0.445760489242956	0.4\\
1.44670523726199	0.442768431711988	0.4\\
1.43458640733652	0.439000163742934	0.4\\
1.42273097486137	0.434470858820916	0.4\\
1.41118667747885	0.429198754858267	0.4\\
1.4	0.423205080756888	0.4\\
1.38921598722678	0.416513970926954	0.4\\
1.37887806257247	0.409152368106167	0.4\\
1.36902785321094	0.401149914870852	0.4\\
1.35970502245874	0.392538834275773	0.4\\
1.35094711006485	0.383353800103258	0.4\\
1.34278938105144	0.373631797244121	0.4\\
1.33526468371403	0.363411972772554	0.4\\
1.328403317353	0.352735478314681	0.4\\
1.32223291026902	0.341645304345482	0.4\\
1.31677830851359	0.330186107081323	0.4\\
1.31206147584282	0.318404028665134	0.4\\
1.3081014052771	0.306346511368286	0.4\\
1.30491404262292	0.294062106557308	0.4\\
1.30251222226472	0.28160027919467	0.4\\
1.30090561548538	0.269011208660837	0.4\\
1.30010069152336	0.256345586699614	0.4\\
1.30010069152336	0.243654413300386	0.4\\
1.30090561548538	0.230988791339163	0.4\\
1.30251222226472	0.21839972080533	0.4\\
1.30491404262292	0.205937893442692	0.4\\
1.3081014052771	0.193653488631714	0.4\\
1.31206147584282	0.181595971334866	0.4\\
1.31677830851359	0.169813892918677	0.4\\
1.32223291026902	0.158354695654518	0.4\\
1.328403317353	0.147264521685319	0.4\\
1.33526468371403	0.136588027227446	0.4\\
1.34278938105144	0.126368202755879	0.4\\
1.35094711006485	0.116646199896742	0.4\\
1.35970502245874	0.107461165724227	0.4\\
1.36902785321094	0.0988500851291483	0.4\\
1.37887806257247	0.0908476318938336	0.4\\
1.38921598722678	0.0834860290730457	0.4\\
1.4	0.0767949192431123	0.4\\
1.41118667747885	0.0708012451417328	0.4\\
1.42273097486137	0.0655291411790837	0.4\\
1.43458640733652	0.0609998362570663	0.4\\
1.44670523726199	0.0572315682880116	0.4\\
1.45903866638696	0.0542395107570443	0.4\\
1.47153703234534	0.0520357116238135	0.4\\
1.48415000862864	0.0506290448096115	0.4\\
1.49682680723304	0.050025174465225	0.4\\
1.50951638316475	0.0502265321633984	0.4\\
1.5221676399802	0.0512323071077492	0.4\\
1.53472963553339	0.0530384493975584	0.4\\
1.54715178710189	0.0556376863352916	0.4\\
1.55938407506565	0.0590195517111852	0.4\\
1.57137724431837	0.0631704279469786	0.4\\
1.58308300260038	0.0680736009290963	0.4\\
1.59445421495454	0.0737093273104835	0.4\\
1.6054450935221	0.0800549140100971	0.4\\
1.61601138191424	0.0870848095899328	0.4\\
1.6261105334169	0.0947707071416486	0.4\\
1.63570188231143	0.103081658268493	0.4\\
1.64474680762101	0.111984197703577	0.4\\
1.6532088886238	0.121442478062692	0.4\\
1.66105405150621	0.131418414189072	0.4\\
1.66825070656624	0.141871836508881	0.4\\
1.67476987541396	0.152760652779906	0.4\\
1.68058530765732	0.164041017582166	0.4\\
1.68567358660321	0.175667508867934	0.4\\
1.69001422354819	0.187593310860303	0.4\\
1.69358974027927	0.199770402563784	0.4\\
1.69638573945254	0.212149751127918	0.4\\
1.69839096256616	0.22468150928525	0.4\\
1.69959733529438	0.237315216068687	0.4\\
1.7	0.25	0.4\\
}--cycle;

\addplot3[area legend, draw=white!20!black, fill=white!20!black, fill opacity=0.3, line width=0.5pt, forget plot]
table[row sep=crcr] {%
x	y	z\\
1.7	0.25	0\\
1.69959733529438	0.262684783931313	0\\
1.69839096256616	0.27531849071475	0\\
1.69638573945254	0.287850248872082	0\\
1.69358974027927	0.300229597436216	0\\
1.69001422354819	0.312406689139697	0\\
1.68567358660321	0.324332491132066	0\\
1.68058530765732	0.335958982417834	0\\
1.67476987541396	0.347239347220094	0\\
1.66825070656624	0.35812816349112	0\\
1.66105405150621	0.368581585810928	0\\
1.6532088886238	0.378557521937308	0\\
1.64474680762101	0.388015802296422	0\\
1.63570188231143	0.396918341731507	0\\
1.6261105334169	0.405229292858351	0\\
1.61601138191424	0.412915190410067	0\\
1.6054450935221	0.419945085989903	0\\
1.59445421495454	0.426290672689516	0\\
1.58308300260038	0.431926399070904	0\\
1.57137724431837	0.436829572053021	0\\
1.55938407506565	0.440980448288815	0\\
1.54715178710189	0.444362313664708	0\\
1.53472963553339	0.446961550602442	0\\
1.5221676399802	0.448767692892251	0\\
1.50951638316475	0.449773467836602	0\\
1.49682680723304	0.449974825534775	0\\
1.48415000862864	0.449370955190388	0\\
1.47153703234534	0.447964288376187	0\\
1.45903866638696	0.445760489242956	0\\
1.44670523726199	0.442768431711988	0\\
1.43458640733652	0.439000163742934	0\\
1.42273097486137	0.434470858820916	0\\
1.41118667747885	0.429198754858267	0\\
1.4	0.423205080756888	0\\
1.38921598722678	0.416513970926954	0\\
1.37887806257247	0.409152368106167	0\\
1.36902785321094	0.401149914870852	0\\
1.35970502245874	0.392538834275773	0\\
1.35094711006485	0.383353800103258	0\\
1.34278938105144	0.373631797244121	0\\
1.33526468371403	0.363411972772554	0\\
1.328403317353	0.352735478314681	0\\
1.32223291026902	0.341645304345482	0\\
1.31677830851359	0.330186107081323	0\\
1.31206147584282	0.318404028665134	0\\
1.3081014052771	0.306346511368286	0\\
1.30491404262292	0.294062106557308	0\\
1.30251222226472	0.28160027919467	0\\
1.30090561548538	0.269011208660837	0\\
1.30010069152336	0.256345586699614	0\\
1.30010069152336	0.243654413300386	0\\
1.30090561548538	0.230988791339163	0\\
1.30251222226472	0.21839972080533	0\\
1.30491404262292	0.205937893442692	0\\
1.3081014052771	0.193653488631714	0\\
1.31206147584282	0.181595971334866	0\\
1.31677830851359	0.169813892918677	0\\
1.32223291026902	0.158354695654518	0\\
1.328403317353	0.147264521685319	0\\
1.33526468371403	0.136588027227446	0\\
1.34278938105144	0.126368202755879	0\\
1.35094711006485	0.116646199896742	0\\
1.35970502245874	0.107461165724227	0\\
1.36902785321094	0.0988500851291483	0\\
1.37887806257247	0.0908476318938336	0\\
1.38921598722678	0.0834860290730457	0\\
1.4	0.0767949192431123	0\\
1.41118667747885	0.0708012451417328	0\\
1.42273097486137	0.0655291411790837	0\\
1.43458640733652	0.0609998362570663	0\\
1.44670523726199	0.0572315682880116	0\\
1.45903866638696	0.0542395107570443	0\\
1.47153703234534	0.0520357116238135	0\\
1.48415000862864	0.0506290448096115	0\\
1.49682680723304	0.050025174465225	0\\
1.50951638316475	0.0502265321633984	0\\
1.5221676399802	0.0512323071077492	0\\
1.53472963553339	0.0530384493975584	0\\
1.54715178710189	0.0556376863352916	0\\
1.55938407506565	0.0590195517111852	0\\
1.57137724431837	0.0631704279469786	0\\
1.58308300260038	0.0680736009290963	0\\
1.59445421495454	0.0737093273104835	0\\
1.6054450935221	0.0800549140100971	0\\
1.61601138191424	0.0870848095899328	0\\
1.6261105334169	0.0947707071416486	0\\
1.63570188231143	0.103081658268493	0\\
1.64474680762101	0.111984197703577	0\\
1.6532088886238	0.121442478062692	0\\
1.66105405150621	0.131418414189072	0\\
1.66825070656624	0.141871836508881	0\\
1.67476987541396	0.152760652779906	0\\
1.68058530765732	0.164041017582166	0\\
1.68567358660321	0.175667508867934	0\\
1.69001422354819	0.187593310860303	0\\
1.69358974027927	0.199770402563784	0\\
1.69638573945254	0.212149751127918	0\\
1.69839096256616	0.22468150928525	0\\
1.69959733529438	0.237315216068687	0\\
1.7	0.25	0\\
}--cycle;

\addplot3 [color=mycolor2, loosely dashed, line cap = round, rounded corners, line width=0.7pt]
 table[row sep=crcr] {%
1	1.2	1\\
1.5714	0.4368	0.4\\
};
 
\addplot3[area legend, line width=1.2pt, line cap = round, draw=mycolor3, fill=mycolor3, fill opacity=0.3, forget plot]
table[row sep=crcr] {%
x	y	z\\
-1.22464679914735e-17	0.2	1.95\\
1.22464679914735e-17	0.6	1.95\\
1.22464679914735e-17	0.6	2.05\\
-1.22464679914735e-17	0.2	2.05\\
}--cycle;
\addplot3 [color=mycolor3, line cap = round, line width=1.2pt, only marks, mark size=0.3pt, mark=*, mark options={solid, mycolor3}]
 table[row sep=crcr] {%
-9.18485099360515e-18	0.25	2\\
-6.56060785257511e-18	0.292857142857143	2\\
-3.93636471154506e-18	0.335714285714286	2\\
-1.31212157051502e-18	0.378571428571429	2\\
1.31212157051502e-18	0.421428571428571	2\\
3.93636471154506e-18	0.464285714285714	2\\
6.56060785257511e-18	0.507142857142857	2\\
9.18485099360515e-18	0.55	2\\
};
 \addplot3 [color=mycolor2, line cap = round, line width=0.7pt]
 table[row sep=crcr] {%
1	1.2	1\\
0	0.4	2\\
};
 \node[above=0.25mm,anchor=center,align=center,font=\footnotesize]
at (axis cs:0,0.4,2.15) {RS $n$};


 \node[anchor=north east,inner sep=0mm,align=center] at (axis cs:0,0.43,1.81) {\scriptsize$\pprs_n\!=\!\!\begin{bmatrix} \xrs_n \\ \yrs_n \\ \zrs_n\end{bmatrix}$};

\node[align=center] at (axis cs:0.6,0.87,2.02) {\scriptsize$\theta_n$};

\addplot3 [ line cap = round, line width=0.5pt]
 table[row sep=crcr] {
0.5	0.4	2\\
0	0.4	2\\
};


  \addplot3[domain=0:-38.7,samples = 80,samples y=0, line width=0.3pt, line cap = round,-latex]
({0.4*cos(-x)},
{0.4+0.4*sin(-x)},
2);

  \addplot3[domain=0:1,samples = 10,samples y=0, line cap = round, line width=0.5pt]
({x*0.5*cos(-321)},
{0.4+x*0.5*sin(-321)},
2);

  \addplot3[domain=1.6793:2,samples = 10,samples y=0, line width=0.5pt, line cap = round, draw = mygray, dashed]
({0.4*cos(-321)},
{0.4+0.4*sin(-321)},
x);

 
\addplot3[area legend, line cap = round, line width=1.2pt, draw=mycolor3, fill=mycolor3, fill opacity=0.3, forget plot]
table[row sep=crcr] {%
x	y	z\\
-1.22464679914735e-17	1	1.95\\
1.22464679914735e-17	1.4	1.95\\
1.22464679914735e-17	1.4	2.05\\
-1.22464679914735e-17	1	2.05\\
}--cycle;
\addplot3 [color=mycolor3, line cap = round, line width=1.2pt, only marks, mark size=0.3pt, mark=*, mark options={solid, mycolor3}]
 table[row sep=crcr] {%
-9.18485099360515e-18	1.05	2\\
-6.56060785257511e-18	1.09285714285714	2\\
-3.93636471154506e-18	1.13571428571429	2\\
-1.31212157051502e-18	1.17857142857143	2\\
1.31212157051502e-18	1.22142857142857	2\\
3.93636471154506e-18	1.26428571428571	2\\
6.56060785257511e-18	1.30714285714286	2\\
9.18485099360515e-18	1.35	2\\
};
 \addplot3 [color=mycolor2, line cap = round, line width=0.7pt]
 table[row sep=crcr] {%
1	1.2	1\\
0	1.2	2\\
};
 \node[align=center]
at (axis cs:0,1.2,2.14) {$\hdots$};
\addplot3 [color=mycolor2, loosely dashed, line cap = round, line width=0.7pt]
 table[row sep=crcr] {%
0.35	1.5	0.35\\
0	1.2	2\\
};
 
\addplot3[area legend, line cap = round, line width=1.2pt, draw=mycolor3, fill=mycolor3, fill opacity=0.3, forget plot]
table[row sep=crcr] {%
x	y	z\\
-1.22464679914735e-17	1.8	1.95\\
1.22464679914735e-17	2.2	1.95\\
1.22464679914735e-17	2.2	2.05\\
-1.22464679914735e-17	1.8	2.05\\
}--cycle;
\addplot3 [color=mycolor3, line cap = round, line width=1.2pt, only marks, mark size=0.3pt, mark=*, mark options={solid, mycolor3}]
 table[row sep=crcr] {%
-9.18485099360515e-18	1.85	2\\
-6.56060785257511e-18	1.89285714285714	2\\
-3.93636471154506e-18	1.93571428571429	2\\
-1.31212157051502e-18	1.97857142857143	2\\
1.31212157051502e-18	2.02142857142857	2\\
3.93636471154506e-18	2.06428571428571	2\\
6.56060785257511e-18	2.10714285714286	2\\
9.18485099360515e-18	2.15	2\\
};
 \addplot3 [color=mycolor2, line cap = round, line width=0.7pt]
 table[row sep=crcr] {%
1	1.2	1\\
0	2	2\\
};
 \node[above=-0.5mm,inner sep=0mm,anchor=south,align=center,font=\footnotesize]
at (axis cs:0,2,2.14) {RS 1};
\addplot3 [color=mycolor2, loosely dashed, line cap = round, line width=0.7pt]
 table[row sep=crcr] {%
0.35	1.5	0.35\\
0	2	2\\
};
 
\addplot3[area legend, line cap = round, line width=1.2pt, draw=mycolor3, fill=mycolor3, fill opacity=0.3, forget plot]
table[row sep=crcr] {%
x	y	z\\
1.4	0	1.95\\
1.8	0	1.95\\
1.8	0	2.05\\
1.4	0	2.05\\
}--cycle;
\addplot3 [color=mycolor3, line cap = round, line width=1.2pt, only marks, mark size=0.3pt, mark=*, mark options={solid, mycolor3}]
 table[row sep=crcr] {%
1.45	0	2\\
1.49285714285714	0	2\\
1.53571428571429	0	2\\
1.57857142857143	0	2\\
1.62142857142857	0	2\\
1.66428571428571	0	2\\
1.70714285714286	0	2\\
1.75	0	2\\
};
 \addplot3 [color=mycolor2, line cap = round, line width=0.7pt]
 table[row sep=crcr] {%
1	1.2	1\\
1.6	0	2\\
};
 \node[below=4mm,align=center,font=\footnotesize]
at (axis cs:1.6,0,2.14) {RS $N$};
\addplot3 [color=mycolor2, loosely dashed, line width=0.7pt, line cap = round]
 table[row sep=crcr] {%
1.5714	0.4368 	0.4\\
1.6	0	2\\
};

\addplot3 [color=mycolor1, line width=1.2pt, line cap = round, only marks, mark size=1.4pt, mark=square, mark options={solid, mycolor1}]
 table[row sep=crcr] {%
1	1.2	1\\
};


\node[align=left,anchor=west,font=\footnotesize] at (axis cs:1,1.25,0.95) {UE \\[-0.5mm] $\pp = [p_x \, p_y \, p_z]^\trp$};



\draw[draw = mygray, solid, line width=0.5pt] (1,1.2,1) -- (1,1.2,0);
\node[align=center] at (axis cs:0.8,0.87,1.35) {\scriptsize$c\,\tau_n$};

\draw[draw = white!15!black, solid, line width=0.5pt] (1.5,0.463,0.41) -- (1.5,0.463,0.025);
\draw[draw = white!15!black, solid, line width=0.5pt] (1.6,0.072,0.41) -- (1.6,0.072,0.01);

 \end{axis}
\end{tikzpicture}%

%% file: Figures/bounds-sweep.tex
%
%
\pgfplotsset{every axis/.append style={
  label style={font=\footnotesize},
  legend style={font=\tiny},
  tick label style={font=\footnotesize},
}}

\definecolor{NCP-M2}{rgb}{0.00000,0.44700,0.74100}%
\definecolor{NCP-M4}{rgb}{0.85000,0.32500,0.09800}%
\definecolor{NCP-M6}{rgb}{0.92900,0.69400,0.12500}%
\definecolor{NCP-M8}{rgb}{0.49400,0.18400,0.55600}%
\definecolor{CP-M2}{rgb}{0.00000,0.44700,0.74100}
\definecolor{CP-M4}{rgb}{0.85000,0.32500,0.09800}
\definecolor{CP-M6}{rgb}{0.92900,0.69400,0.12500}
\definecolor{CP-M8}{rgb}{0.49400,0.18400,0.55600}
\begin{tikzpicture}[spy using outlines={circle, magnification=2000, size=1cm, connect spies}]

\begin{axis}[%
width=\plotWidth,
height=\plotHeight,
at={(0\plotWidth,0\plotWidth)},
scale only axis,
xmode=log,
xmin=1000000,
xmax=1000000000,
xminorticks=true,
xlabel={$B$ in Hz},
xlabel style={yshift=0.2cm},
ymode=log,
ymin=0.001,
ymax=1,
yminorticks=true,
ylabel={$\peb$ in m},
ylabel style={yshift=-0.2cm},
axis background/.style={fill=white},
xmajorgrids,
xminorgrids,
ymajorgrids,
yminorgrids,
grid style=dotted,
legend style={at={(0.02,0.4)}, anchor=west, legend cell align=left, align=left, draw=white!15!black,legend columns=2}
]
\addplot [color=NCP-M2,mark=triangle,mark options={solid}, line width=1.25pt, line cap = round]
  table[row sep=crcr]{%
1000000	0.544344155853623\\
2000000	0.542877634045264\\
4000000	0.538797612710415\\
10000000	0.522574805367939\\
20000000	0.489333486803425\\
40000000	0.423610452635098\\
100000000	0.288645135681258\\
200000000	0.164622434077007\\
400000000	0.0681830227913661\\
1000000000	0.0265086734391372\\
};
\addlegendentry{NCP ($M=2$)}

\addplot [color=NCP-M4,mark=triangle,mark options={solid,rotate=90}, line width=1.25pt, line cap = round]
  table[row sep=crcr]{%
1000000	0.244172174334094\\
2000000	0.24403930669856\\
4000000	0.243664880448553\\
10000000	0.242103863446712\\
20000000	0.23850784003465\\
40000000	0.229447731113291\\
100000000	0.197919685044466\\
200000000	0.140562001407277\\
400000000	0.0660855692901826\\
1000000000	0.0263800732459715\\
};
\addlegendentry{NCP ($M=4$)}

\addplot [color=NCP-M6,mark=square,mark options={solid}, line width=1.25pt, line cap = round]
  table[row sep=crcr]{%
1000000	0.159917234594042\\
2000000	0.15987988752285\\
4000000	0.159774445326652\\
10000000	0.15933170448521\\
20000000	0.158292258365614\\
40000000	0.155548768077082\\
100000000	0.144469926838697\\
200000000	0.116871494952255\\
400000000	0.0629963854706102\\
1000000000	0.0261700206531866\\
};
\addlegendentry{NCP ($M=6$)}

\addplot [color=NCP-M8,mark=o,mark options={solid}, line width=1.25pt, line cap = round]
  table[row sep=crcr]{%
1000000	0.119212433899954\\
2000000	0.119196959438108\\
4000000	0.119153243022209\\
10000000	0.118969242055698\\
20000000	0.118534449737068\\
40000000	0.117367827608598\\
100000000	0.112367785828886\\
200000000	0.0977497099593659\\
400000000	0.0593376474750853\\
1000000000	0.0258845583837769\\
};
\addlegendentry{NCP ($M=8$)}

\addplot [densely dashed,mark=o,mark options={solid},color=black, line width=1.25pt, line cap = round]
  table[row sep=crcr]{%
1000000	0.00216749756683927\\
2000000	0.00216719112315539\\
4000000	0.00216657842173784\\
10000000	0.00216474208687027\\
20000000	0.00216168768065928\\
40000000	0.0021556028152353\\
100000000	0.00213754110842832\\
200000000	0.00210803597940298\\
400000000	0.00205085793197155\\
1000000000	0.00189443408507189\\
};
\addlegendentry{CP ($M=2\dots8$)}

\end{axis}

\end{tikzpicture}%

%% file: Figures/RMSE_PEB.tex
%
%
\pgfplotsset{every axis/.append style={
  label style={font=\footnotesize},
  legend style={font=\tiny},
  tick label style={font=\footnotesize},
}}

\definecolor{mycolor1}{rgb}{0.00000,0.44700,0.74100}%
\definecolor{mycolor2}{rgb}{0.85098,0.32549,0.09804}%
\definecolor{mycolor3}{rgb}{0.46600,0.67400,0.18800}%
\definecolor{mycolor4}{rgb}{0.92900,0.69400,0.12500}%
\begin{tikzpicture}

\begin{axis}[%
width=\plotWidth,
height=\plotHeight,
at={(0\plotWidth,0\plotWidth)},
scale only axis,
xmin=0,
xmax=25,
xlabel={$\overline{\mathrm{SDNR}}$ in dB},
ymode=log,
ymin=0.0007,
ymax=5,
yminorticks=true,
ylabel={RMSE in m},
xlabel style={yshift=0.2cm},
ylabel style={yshift=-0.2cm},
axis background/.style={fill=white},
xmajorgrids,
ymajorgrids,
yminorgrids,
grid style=dotted,
legend style={at={(1,1)}, anchor=north east, legend cell align=left, align=left, fill=white, draw=white!15!black,legend columns=3}
]

\addplot [color=mycolor4, line width=1.25pt, mark size=2pt, mark=square, mark options={solid, mycolor4}]
  table[row sep=crcr]{%
0	2.82842712474619\\
5	1.58284271247462\\
10	0.372052057945166\\
15	0.011402051139295\\
20	0.001602284270196\\
25	0.000713947416988668\\
};
\addlegendentry{ML}

\addplot [color=mycolor2, line width=1.25pt, mark size=2pt, mark=diamond, mark options={solid, mycolor2}]
  table[row sep=crcr]{%
0	2.82842712474619\\
5	1.82842712474619\\
10	0.715562951706863\\
15	0.311853768147439\\
20	0.102356853057247\\
25	0.05320950246766\\
};
\addlegendentry{ML-NCP}

\addplot [color=mycolor1, line width=1.25pt, mark size=2.0pt, mark=o, mark options={solid, mycolor1}]
  table[row sep=crcr]{%
0	3.28284271247462\\
5	2.285958\\
10	1.08174335110197\\
15	0.62925867293776\\
20	0.190901742920564\\
25	0.148698763819638\\
};
\addlegendentry{ILS}

\addplot [color=black, densely dashed, line width=1.5pt]
  table[row sep=crcr]{%
0	0.0125324086495412\\
5	0.00704749128780999\\
10	0.00396309559005451\\
15	0.00222861242596728\\
20	0.00125324086495412\\
25	0.000704749128781\\
};
\addlegendentry{PEB}

\addplot [color=black, dashdotted, line width=1.5pt]
  table[row sep=crcr]{%
0	0.939784263719636\\
5	0.528479528253195\\
10	0.297185878253664\\
15	0.167119900605051\\
20	0.0939784263718733\\
25	0.0528479528253056\\
};
\addlegendentry{PEB-NCP}

\end{axis}

\end{tikzpicture}%

%% file: Figures/RMSE_CEB.tex
%
%

\pgfplotsset{every axis/.append style={
  label style={font=\footnotesize},
  legend style={font=\tiny},
  tick label style={font=\footnotesize},
}}

\definecolor{mycolor1}{rgb}{0.00000,0.44700,0.74100}%
\definecolor{mycolor2}{rgb}{0.85098,0.32549,0.09804}%
\definecolor{mycolor3}{rgb}{0.46600,0.67400,0.18800}%
\definecolor{mycolor4}{rgb}{0.92900,0.69400,0.12500}%
\begin{tikzpicture}

\begin{axis}[%
width=\plotWidth,
height=\plotHeight,
at={(0\plotWidth,0\plotWidth)},
scale only axis,
xmin=0,
xmax=25,
xlabel={$\overline{\mathrm{SDNR}}$ in dB},
ymode=log,
ymin=0.1,
ymax=500,
yminorticks=true,
ylabel={RMSE in ns},
axis background/.style={fill=white},
xmajorgrids,
ymajorgrids,
yminorgrids,
xlabel style={yshift=0.2cm},
ylabel style={yshift=-0.2cm},
grid style=dotted,
legend style={at={(1,1)}, anchor=north east, legend cell align=left, align=left, draw=white!15!black,legend columns=3}
]

\addplot [color=mycolor4, line width=1.25pt, mark size=2pt, mark=square, mark options={solid, mycolor4}]
  table[row sep=crcr]{%
0	98.8945967\\
5	56.5824915824916\\
10	12\\
15	1\\
20	0.254621319963631\\
25	0.123938411695197\\
};
\addlegendentry{ML}

\addplot [color=mycolor2, line width=1.25pt, mark size=2pt, mark=diamond, mark options={solid, mycolor2}]
  table[row sep=crcr]{%
0	126.666666666667\\
5	86.7849510705\\
10	32.974893898\\
15	11.5360382111434\\
20	3.24621319963631\\
25	1.62384116951966\\
};
\addlegendentry{ML-NCP}

\addplot [color=mycolor1, line width=1.25pt, mark size=2.0pt, mark=o, mark options={solid, mycolor1}]
  table[row sep=crcr]{%
0	166.6667\\
5	126.4\\
10	78.1156\\
15	34.536\\
20	7.3462\\
25	3.2384\\
};
\addlegendentry{ILS}

\addplot [color=black, densely dashed, line width=1.5pt]
  table[row sep=crcr]{%
0	2.20327297345358\\
5	1.23899144364767\\
10	0.69673609032029\\
15	0.391803496338713\\
20	0.220327297345345\\
25	0.123899144364776\\
};
\addlegendentry{CEB}

\addplot [color=black, dashdotted, line width=1.5pt]
  table[row sep=crcr]{%
0	28.6365285666727\\
5	16.1035034242149\\
10	9.05566545659043\\
15	5.09237491382638\\
20	2.86365285696881\\
25	1.61035034284843\\
};
\addlegendentry{CEB-NCP}

\end{axis}

\end{tikzpicture}%